\documentclass[twocolumn,times,tighten,twocolappendix]{aastex63}
\usepackage{graphicx,amssymb,amsmath,amsthm,amsfonts,epsfig}
\newcommand{\fGW}{f_{\text{GW}}}
\newcommand{\nus}{\nu_{\star}}
\newcommand{\Ms}{M_{\star}}
\newcommand{\Rs}{R_{\star}}
\newcommand{\Bs}{B_{\star}}

\begin{document}

\title{Wave-optical effects in the microlensing of continuous gravitational waves by star clusters}

\author{Arthur G. Suvorov} \thanks{arthur.suvorov@manlyastrophysics.org}
\affil{Manly Astrophysics, 15/41-42 East Esplanade, Manly, NSW 2095, Australia}

\begin{abstract}
\noindent{Rapidly rotating neutron stars are promising sources for existing and upcoming gravitational-wave interferometers. While relatively dim, these systems are expected to emit continuously, allowing for signal to be accumulated through persistent monitoring over year-long timescales. If, at some point during the observational window, the source comes to lie behind a dense collection of stars, transient gravitational lensing may occur. Such events, though rare, would modulate the waveform, induce phase drifts, and ultimately affect parameter inferences concerning the nuclear equation of state and/or magnetic field structure of the neutron star. Importantly, the radiation wavelength will typically exceed the Schwarzschild radius of the individual perturbers in this scenario, implying that (micro-)lensing occurs in the diffractive regime where geometric optics does not apply. In this paper, we make use of numerical tools that borrow from Picard-Lefschetz theory to efficiently evaluate the relevant Fresnel-Kirchhoff integrals for $n \gtrsim 10^{2}$ microlenses. Modulated strain profiles are constructed both in general and for particular neutron star trajectories relative to some simulated macrolenses.
}
\end{abstract}


\section{Introduction} %\label{sec:intro}
%, most notably the advanced Laser Interferometer Gravitational-Wave Observatory (aLIGO)
%\citep{bild98,git19} \cite[though see also][] {pat12,glamp21}. 

Certain classes of neutron stars are expected to be excellent sources of continuous gravitational waves (GWs). Anomalous X-ray pulsars and soft gamma repeaters, now widely recognised as neutron stars of the highly-magnetised variety \cite[`magnetars',][]{td92}, may be deformed by magnetic stresses to the point that the resulting GW luminosity could be detected by currently operating ground-based interferometers \citep{cf53,goos72,mast11}. Neutron stars accreting through Roche-lobe overflow are another strong candidate, since there remains an observational puzzle as to why their spin frequencies seem to be capped at $\lesssim 700$ Hz \citep{pat17}; such systems would be expected to spin-up indefinitely unless stalled by a sufficiently large, spin-dependent counter-torque. It has been argued that GW radiation-reaction may be the key agent that limits the rotational growth \cite[\citealt{bild98,git19}; though see also][]{pat12,glamp21}. Regardless, because the GW strain scales quadratically with the spin frequency, some of the most promising sources for long-term emissions are those with millisecond periods, where the resulting GW frequency, $\fGW$, lies in the $\sim$kHz band \citep{thorne80,ligo21}.

For radiation at these frequencies, gravitational or otherwise, wave-optical effects are expected to come into play when encountering solar-mass bodies along or near the line of sight \citep{oh74,nak99,mac04}. Diffractive effects in particular are important for $M_{L} \lesssim 10^{2} M_{\odot} (\fGW / \text{kHz})^{-1}$ \citep{tak03,tak05}, when the wavelength of the source exceeds the Schwarzschild radius of the (micro-)lens. For macrolenses consisting of $n \gtrsim 10^{2}$ stars, we then enter into an intermediate arena between the heavily diffracted and eikonal regimes, where the overall amplifications may be non-negligible and `beat' patterns can emerge at the interferometer due to time delays \citep{christ18,jung19,cheung21}. Although impressive advances have been made in the numerical implementation of ray-shooting codes in geometric optics \citep{lew10,lew20}, such tools are not applicable in this case. Wave-optical lensing for \emph{continuous} GWs may be especially impactful because detections would likely take many months of persistent monitoring using phase-coherent strategies \citep{lasky15,derg21,suv21,sold21}, and the line of sight may cross a number of interference fringes during this time. 

%MENTION SOMEWHERE HERE ITS RARE.

In this respect, \cite{liao19} have demonstrated that diffraction and interference effects may, albeit rarely, show up in continuous GW signals, and that the amplitude and phase modulations arising due to lensing can non-negligibly affect parameter estimation \cite[see also][]{dep01,diego19,marc20,meena20,mishra21}. However, these authors concentrated on the case of a single point-mass lens, where a closed-form expression for the lensing flux is available \citep{nak99}, thereby allowing for analytically-tractable computations. While lensing by multiple stars is highly unlikely for any given \emph{Galactic} source \citep{pac86b,jow20}, it is nevertheless worthwhile to re-examine the situation for different macro configurations. Generally speaking, the main challenge in performing legitimately wave-optical calculations is the oscillatory nature of the relevant Fresnel-Kirchhoff diffraction integral \citep{peters74}; given a phase profile $\varphi$, the physical optics calculation involves integrating the exponential $e^{i \varphi}$ over the aperture, which is infinitely oscillatory owing to Euler's formula. \cite{feld19}, following a mathematical program outlined by \cite{witten11}, have recently developed a method based on the application of Picard-Lefschetz (PL) theory that is useful in this regard \cite[see also][for a catalogue of other methods]{guo20}.

In essence, the PL calculation involves analytically continuing the integrand into the complex plane. One then builds a set of special contours (`Lefschetz thimbles') which ultimately form a closed loop, so that Cauchy's theorem may be applied. The actual Fresnel-Kirchhoff integral of interest can then be evaluated by calculating instead some simpler, non-oscillatory integrals. Each Lefschetz thimble is associated to a point of stationary phase (i.e., an image) of the original integrand, thereby connecting back to the more familiar geometric optics calculation. The main novelty of this paper is that, by adopting the PL methodology described by \cite{feld19,feld20a,feld20b}, we are able to perform wave-optical calculations for $\sim$kHz GWs lensed by clusters consisting of $\gtrsim10^{2}$ stars. Such a scenario may be relevant when observing neutron stars located behind particularly dense regions of the Galaxy with the next generation of detectors \citep{pac86b,liao19}. GWs, unlike light, also tend to propagate through matter without scattering and absorption, so that lensing remains nominally important even through regions that are opaque in the optical. %RE? %(or other electromagnetic) bands.

This paper is organised as follows. In Section 2 we review the theory of continuous GW generation by deformed neutron stars, outlining the potential impact of wave-optical lensing. Section 3 is then devoted to the derivation of the relevant Fresnel-Kirchhoff integral in the thin-lens approximation, and the setting up of microlens distributions. The numerical techniques, based on PL theory, are described in Section 4 {(further tests and worked examples are given in the Appendices)}, with the results given in Section 5. Some discussion is presented in Section 6.

%\citep{diego19,mishra21,cheung21}

\section{Continuous gravitational waves from neutron stars}

GWs emitted by a non-axisymmetric system are polarised according to how momentum (current multipoles) and energy (mass multipoles) are distributed within the host body \citep{thorne80}. For rapidly rotating neutron stars, several mechanisms can organically induce large momentum or energy fluxes within the stellar interior. For instance, a sufficiently strong magnetic field introduces density asymmetries within the core and outer layers \citep{cf53,goos72}, generating a time-dependent mass quadrupole moment, conventionally written $\ddot{I}_{22} \propto \nus^2 e^{2 i \nus t} I_{0} \varepsilon$ for spin frequency $\nus$, moment of inertia $I_{0}$, and triaxial ellipticity $\varepsilon$ \citep{lasky15}. Mode oscillations, possibly driven to large amplitudes through secular instabilities \citep{and99}, are another often-considered possibility for exciting multipoles of either the current or mass variety \citep{owen10,stergbook}. For concreteness we focus on magnetic deformations, where GWs are emitted at twice the rotational frequency, $\fGW = 2 \nus$, and carry an intrinsic amplitude of \cite[e.g.,][]{lasky15}
\begin{equation} \label{eq:massquadamp}
\begin{aligned}
h_{0} =& \frac {4 \pi^2 G \varepsilon I_{0} \fGW^2} {c^4 D_{\text{OS}}} \\
\approx& \, 1.1 \times 10^{-27} \left( \frac {\varepsilon} {10^{-8}} \right) \left( \frac{\nus} {500 \text{ Hz}} \right)^{2} \left( \frac {10 \text{ kpc}} {D_{\text{OS}}} \right),
\end{aligned}
\end{equation}
as measured by an observer a distance $D_{\text{OS}}$ from the source. 

For a neutron star consisting of normal $npe$ matter, the ellipticity is roughly equal to the ratio of magnetic energy to gravitational binding energy. In terms of a characteristic field strength $\Bs$, one can estimate \citep{mast11}
%7.4
\begin{equation} \label{eq:massquad}
\varepsilon \approx 4 \times 10^{-8} \left( \frac{\Bs}{10^{14} \text{ G}}\right)^2 \left( \frac {\Rs} {10^{6} \text{ cm}} \right)^{4}  \left( \frac {1.4 M_{\odot}} {\Ms} \right)^{2},
\end{equation}
assuming a purely poloidal and dipolar configuration on top of a hydrostatic Tolman-VII density profile. The inclusion of higher multipoles, a toroidal field, or employing a different equation of state can potentially lead to order-of-magnitude adjustments within expression \eqref{eq:massquad} \citep{dos09,cio09}. Additionally, for a star whose core contains superconducting protons, $\varepsilon$ is amplified by a factor $\sim H_{c1}/\Bs$ \citep{cutler02}, where $H_{c1} \lesssim 10^{16} \text{ G}$ represents the microscopic critical field strength, the exact value of which is determined by the London penetration depth, amongst other physical factors \citep{glamp11,lander13}. Stars with surface fields $B_{\star} \lesssim 10^{12} \text{ G}$ may therefore still permit strains of order $h_{0} \gtrsim 10^{-27}$ if their cores are superconducting or if they harbour a dominant toroidal field \citep{suv21}. Even in the restricted context of magnetic deformations, it is clear therefore that a detection of continuous GWs from a localised source, which would effectively measure $\varepsilon$ within some tolerance, can yield a significant amount of information about stellar structure.

%\begin{equation}
%\varepsilon = 3.4 \times 10^{-8} \left( \frac{\Bs}{10^{12} \text{ G}}\right) \left( \frac{\H_{c1}}{10^{16} \text{ G}}\right) 
%\end{equation}

\subsection{Detectability and relative motion}

The characteristic strains \eqref{eq:massquadamp} are orders of magnitude lower than those due to the violent merger events that have thus far been detected. Persistent emissions, associated with magnetically (or otherwise) deformed neutron stars, have the advantage however that signal can be accrued over many cycles. In particular, if the star houses a mass or current quadrupole moment with a lifetime that exceeds the observational window $T_{\text{obs}}$, continual monitoring leads to an increase in detector sensitivity. In a fully-coherent search, a ground-based interferometer can detect a signal of amplitude 
\begin{equation} \label{eq:h0thresh}
h_{0} \approx 11.4 \sqrt{ S_{n} / T_{\text{obs}}}
\end{equation}
with $90\%$ confidence \citep{watts08}, where $S_{n}$ is the noise power spectral density of the detector. Although not shown here \cite[see, for example, Figure 5 in][]{sold21}, it is likely that observations spanning at least a year would be necessary to detect continuous GWs from many of the known sources with existing instruments \cite[see also][]{lasky15,suv21}.

Suppose however that the GWs from the system were lensed en route to the detector. In the case of burst-like signals associated with mergers, for example, where the bulk of the measurable GW luminosity is emitted within a time window spanning a few seconds, relative motion between the lens and source is negligible. Still, wave-like effects are likely to be important here because the source frequency sweeps (`chirps') through a wide band, and the lensing-induced amplification is an oscillatory function of $\fGW$ \citep{nak99,tak03,christ18}.

By contrast, while continuous GWs are expected to be roughly monochromatic (though see below), the relative motion between the lens and source cannot be ignored when considering observational windows spanning some months. For a phase-coherent search lasting $\gtrsim$ one year, the neutron star would have travelled a (relative) distance of $\sim 10^{15} \times (v / 300 \text{ km s}^{-1}) (T_{\text{obs}}/ \text{ yr} )$ cm, where $v = 300 \text{ km s}^{-1}$ is a typical transverse velocity for a millisecond pulsar \citep{hobbs06}. This distance, while negligible compared to $D_{\text{OS}} \sim 10$ kpc, comfortably exceeds the Einstein radius associated with a solar-mass microlens, viz.
\begin{equation} \label{eq:einrad}
\begin{aligned}
R_{E} &\approx \sqrt{ \frac{ 4 G M_{L}} {c^2} \frac{ D_{\text{OL}} D_{\text{LS}}} {D_{\text{OS}}} } \\
&= 6.7 \times 10^{13} \left( \frac {M_{L}} {M_{\odot}} \right)^{1/2}   \text{ cm},
\end{aligned}
\end{equation}
for a lens located $D_{\text{OL}} = 5$ kpc from the observer and a further\footnote{We ignore cosmological corrections to all distance quantities since $z \ll 1$ for the sources considered here.} $D_{\text{LS}} = 5$ kpc from the source. In rare cases, the emitter may thus cross multiple Einstein rings over a long $T_{\text{obs}}$ when located behind particularly dense regions of the Galaxy \cite[see][for some rate estimates]{dep01,jow20}. Crossing interference fringes will lead to modulations of the GW signal, and could noticeably affect parameter estimation in the event of a detection, depending on the location of the neutron star and its environs.

Moreover, wave-optical lensing will generally cause the phase of the signal to drift over time. This may be problematic for phase-coherent GW searches, as matched filtering generally requires the (noisy) detector output, multiplied by a template waveform, to remain in phase with the signal to within $\lesssim$ 1 rad \cite[e.g.,][]{jones04}. This problem is well known in the case of searches directed at sources within active binaries, where the GW frequency can drift due to accretion-induced spin evolution and it is necessary to analyse the signal semi-coherently over segments shorter than the full $T_{\text{obs}}$ \citep{dre18}. The spins of isolated neutron stars may also wander over $\sim$ year-long timescales for a variety of reasons (e.g., glitches); see \cite{suva16} for a discussion. One aspect we can explore with wave-optical calculations is, for a given macrolens, the maximum interval over which matched filtering can be reliably applied (see Sec. 5).

 %The ellipticity of the system may also be time-dependent in this case, while in the lensing it may only appear to be the case because 

%\footnote{Note however that neutron star spins may wander over $\sim$ year-long timescales for a variety of reasons (e.g., glitches). We will however ignore such complications; see \cite{suva16} for a discussion.}
%One of the aspects we explore here then is  maximum interval over which the signal can be reliably analysed with fully coherent methods, whic

%even if the effect of lensing is equivalent to a time-dependent amplification of $\varepsilon$, the braking index of the pulsar will not be likewise changing. In particular, since the GW torque scales as the square of the \emph{unlensed} ellipticity, independent measurements of the spin evolution of the system can, in principle, be used to discern the effects of lensing.

\section{Wave-optical microlensing of gravitational waves}

Here we briefly review the wave-optical theory of gravitational (micro-)lensing of GWs emitted by a point source, closely following \cite{tak03} and \cite{mac04}. Working within the weak lensing regime, we adopt the `Newtonian' spacetime metric
\begin{equation}
  ds^2 = g_{\mu\nu}^{(L)}dx^\mu dx^\nu = - \left( 1+2U \right) dt^2 + \left( 1-2U \right)
 d \boldsymbol{x}^2 ,
\label{eq:metric}  
\end{equation}
where $U(\boldsymbol{x}) \ll 1$ denotes the gravitational potential associated with the macrolens and we have temporarily set the speed of light, $c$, to unity. At the linear level, the macro potential $U$ is simply the superposition of potentials $U_{k}$ associated with each microlensing body $k$, i.e., $U = \sum_{k \leq n} U_{k}$ for $n$ microlenses. GWs emitted by the source are described by a vacuum perturbation $h$ on top of the lensing background, viz. $g_{\mu \nu} = g_{\mu\nu}^{({L})} + h_{\mu \nu}$, which satisfies \citep{isaac68}
\begin{equation} \label{eq:pert}
0 = \nabla_{\alpha} \nabla^{\alpha} h_{\mu\nu} +2R^{(L)}_{\alpha\mu\beta\nu} h^{\alpha\beta} + \mathcal{O}(h^2),
\end{equation}
where $R^{(L)}_{\alpha \beta \mu \nu}$ is the Riemann tensor associated with $g_{\mu \nu}^{(L)}$ and we have employed the Lorentz gauge ($\nabla_{\nu} h^\nu_{\mu}=0$ and ${h^\mu}_{\mu}=0$). For the cases considered here, the GW wavelength is tiny relative to the typical radius of curvature associated with the background, and we can safely drop the Riemann tensor term in equation \eqref{eq:pert}. In this instance, the equations of motion for each component of $h$ are individually equivalent to a Klein-Gordon equation for a scalar field $\phi(t,\boldsymbol{x})$ \citep{peters74}. The leading-order problem thus reduces to finding solutions to
\begin{equation}
0 = \left( \nabla^2 + \omega^2 \right) \phi - 4 \omega^2 U \phi,
\label{eq:scalar}
\end{equation}
where, through a slight abuse of notation, we have taken out the time dependence via $\phi = e^{i \omega t} \phi$ with $\omega = 2 \pi \fGW$. Finally, equation \eqref{eq:scalar} can be solved using Kirchhoff's theorem, thereby defining the Fresnel-Kirchhoff diffraction integral associated with the wave optics of GW lensing \citep{tak05},
\begin{equation} \label{eq:truefk}
\phi(\boldsymbol{x}) = \phi^{(0)}(\boldsymbol{x}) - \frac {\omega^2} {\pi} \int d^{3} \boldsymbol{x}' \frac {e^{i \omega |\boldsymbol{x} - \boldsymbol{x}'|}} {|\boldsymbol{x} - \boldsymbol{x}'|} U(\boldsymbol{x}') \phi^{(0)}(\boldsymbol{x}'),
\end{equation}
where $\phi^{(0)}$ solves the homogeneous ($U=0$) version of equation \eqref{eq:scalar}. Expression \eqref{eq:truefk} can also be obtained from a path integral \citep{nak99}, in line with the expectation that the wave-optical equations can be derived from quantum-mechanical arguments; see also \cite{feld19} for a derivation in the case of electromagnetic radiation, where the resulting equation(s) are practically identical.

% [see \cite{nak99} for details and \cite{tak05} for the relevant equations for geometrically-thick lenses]
\subsection{Thin-lens approximation}

As it stands, the Fresnel-Kirchhoff integral \eqref{eq:truefk} contains a non-local Green's function. We introduce the thin-lens approximation to reduce the dimensionality of the problem and to eliminate the unwieldy denominator term; this amounts to projecting each microlens onto a single 2-dimensional screen, the mathematical details of which can be found, for example, in \cite{tak05}. The end result is that the amplification factor, $F = \phi/\phi^{(0)}$, can be written as [see equation (2.11) in \cite{nak99}]
\begin{equation}
F(\boldsymbol{x}_{s}) = \frac {4 G M_{L}} {c^3} \frac {\fGW} {i} \int d^2 \boldsymbol{x} \exp\left[2\pi i \fGW t_d(\boldsymbol{x},\boldsymbol{x}_{s})\right],
\label{eq:diffint} 
\end{equation}
where $\boldsymbol{x}$ (lens-plane-projected coordinates) and $\boldsymbol{x}_{s}$ (source-plane-projected coordinates) are expressed in units of 
\begin{equation} \label{eq:units}
\xi_{0} = R_{E} \,\,\,\, \text{and} \,\,\,\, \eta_{0} = \frac{D_{\text{OS}}} {D_{\text{OL}}} R_{E},
\end{equation}
respectively, and the Einstein radius $R_{E}$ is defined in expression \eqref{eq:einrad}. For a lens that is equidistant between the source and the observer, we have $\eta_{0} = 2 \xi_{0}$. The (normalised) time delay $t_{d}$, up to a constant factor, is a sum of the geometric and Shapiro delays,
\begin{equation} \label{eq:timedelay}
t_d(\boldsymbol{x},\boldsymbol{x}_{s})= \frac{4 G M_{L}} {c^3} \left[ \frac{1}{2}|\boldsymbol{x}-\boldsymbol{x}_{s}|^2+\psi(\boldsymbol{x}) \right],
\end{equation}
where $\psi(\boldsymbol{x})$ is the dimensionless deflection potential (i.e., the projection of $U$ onto the 2D lens screen). For a collection of $n$ point lenses, we have
\begin{equation} \label{eq:psipot}
\psi(\boldsymbol{x}) = -\displaystyle{\underset{k \leq n}{\sum}} \left( \frac {M_{k}} {M_{L}} \right) \log \sqrt{ \left( x - x_{k} \right)^2 + \left( y - y_{k} \right)^2 },
\end{equation}
where the bodies of mass $M_{k}$ are located at $(x_{k},y_{k})$ on the lens plane. 

We close this section by noting that in microlensing calculations, one generally also includes convergence ($\kappa$) and shear $(\gamma)$ components related to `off-screen' elements within the time-delay function \eqref{eq:timedelay} \cite[see, e.g.,][]{pac86,meena20,lew20}. The former optical scalar accounts for magnifications due to a smooth mass component (e.g., a background Galactic contribution) while the latter is induced by large-scale anisotropies (e.g., tidal distortions). While the formalism presented in the next section can also be used to study cases with non-zero convergence or shear, we defer such a calculation to a future work.

\subsection{Star clusters}

%As such, this fraction can be roughly estimated as the number of neutron stars with millisecond periods relative to the overall population.
%More distant sources (e.g., behind the bulge) have, in general, a higher chance of being lensed \citep{jow20}, though are also harder to detect because the strain $h_{0} \propto 1/D_{OS}$.

For sources out to $\gtrsim 10$~kpc, the likelihood that any emitted GWs non-negligibly interact with the gravitational field of a perturber en route to Earth is relatively low: the lensing probability by stars for a source within the bulge may reach a few times $10^{-6}$ \citep{pac86b,dep01}. If, however, a non-negligible fraction\footnote{\cite{woan18} have provided population-based evidence that all millisecond pulsars house a \emph{minimum} ellipticity of $\varepsilon \sim 10^{-9}$, comparable to expression \eqref{eq:massquad}, a fraction $f_{l}$ of which may be observable in GWs. Although dependent on recycling and star formation assumptions, \cite{zhu15} estimate that the millisecond neutron star birth rate is $\sim 10^{-4} \text{ yr}^{-1}$ from the combined core collapse and accretion-induced collapse channels. A rough estimate for the number of observable sources is then $\sim 10^{6} \times f_{l}$, further multiplied by the fraction of the source's lifetime \cite[$\lesssim$Gyr,][]{zhu15} relative to the age of the Milky Way.} \cite[$\approx 10^{-5}$,][]{liao19} of the $\sim 10^{9}$ neutron stars within the bulge emit appreciable GWs, microlensing events may conceivably be observed by the next generation of interferometers over long $T_{\text{obs}}$. Furthermore, \cite{ras21} have recently suggested that for neutron stars in the globular clusters 47 Tuc and M22, `self-lensing' (i.e., lensing by a fellow member of the cluster) rates may be as high as $2 \times 10^{-3} \text{ yr}^{-1}$ and $4 \times 10^{-5} \text{ yr}^{-1}$, respectively. The single-lens scenario has been considered in detail by \cite{liao19}, who made use of the fact that the Fresnel-Kirchhoff integral can be evaluated analytically in the case of an individual point-mass lens {(see Appendix B)}. Here, we generalise their scenario by considering instead macrolenses consisting of $n \gtrsim 10^{2}$ stars at various distances from the line of sight.

%A conceptually straightforward extension to this idea is to consider a macrolens consisting of multiple objects of varying distance relative to the line of sight.

%However, if the source resides within or behind a particularly dense section of the Galaxy (e.g., a globular cluster or in the Galactic bulge), multiple microlensing events may take place, leading to a larger, collective effect \citep{christ18}. For example, the heart of the globular cluster 47 Tucanae lies $\sim 4$ kpc from Earth, boasting a large ($\gtrsim 10^{5}$) number of stars and at least 25 millisecond pulsars \citep{fre17,rid21}. Although improbable for any given source, it is not difficult to imagine a scenario where GWs from a set of sources with high velocities, originating in dense clusters like 47 Tuc, are lensed by several stars within a year-long observational run. %While improbable for any given source, 

% especially for accreting sources where the companion may periodically eclipse the waves \citep{marc20}.
%The impact of any individual encounter depends on the relative impact parameter

%While it is improbable that non-negligible lensing occurs for any given source, 
 %We aim to address the question of how the GW signal is modulated in such a scenario.
 
 \begin{figure}[h]
  \includegraphics[width=0.49\textwidth]{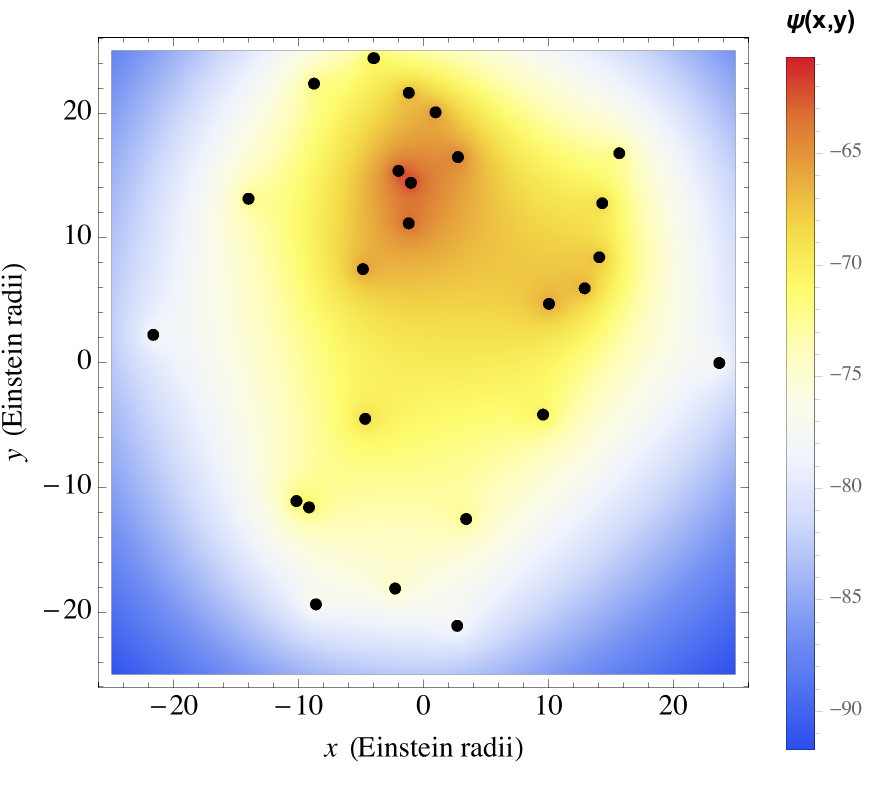}
  \caption{Distribution of $n=25$ microlenses (black dots) over the lens plane with a total area $A = 2500 R_{E}^2$. The colour scale depicts the potential $\psi$ from equation \eqref{eq:psipot}, with redder shades indicating a stronger (i.e., less negative) value. 
}
  \label{fig:grav25}
\end{figure}

\begin{figure}[h]
  \includegraphics[width=0.49\textwidth]{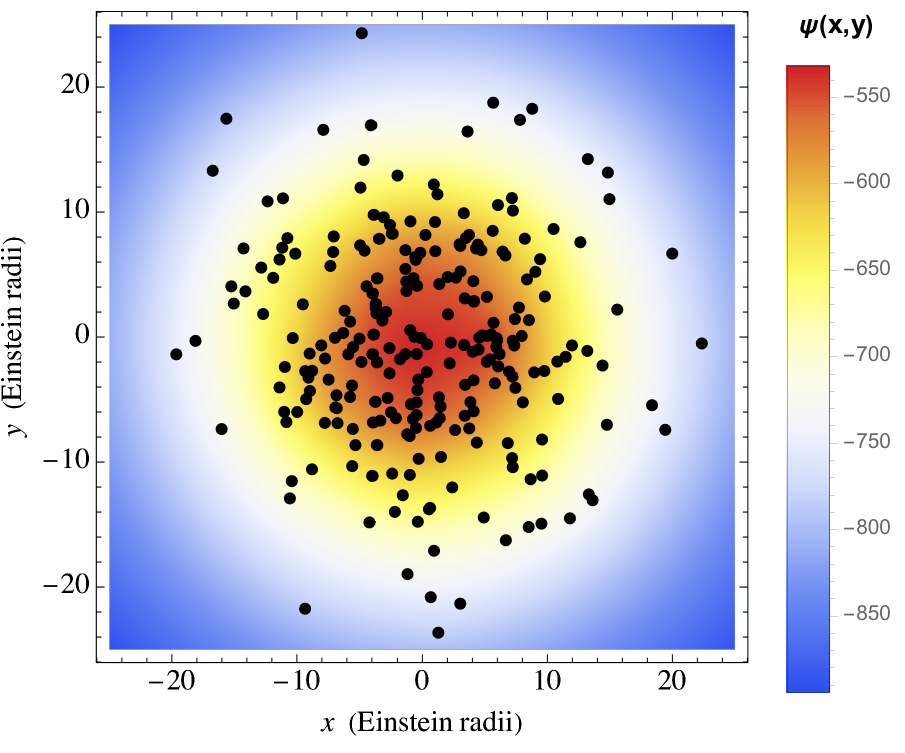}
  \caption{Similar to Fig. \ref{fig:grav25}, though for $n=250$ microlenses distributed within the same area. 
}
  \label{fig:grav250}
\end{figure}

For concreteness, we consider two distinct cluster distributions in this work. Defining an `optical depth', $\tau$, as the ratio of the area covered by the individual Einstein rings to a fiducial screen area $A$, expressed in units of $R_{E}$, we have $\tau \approx n \pi / A$. If we consider a $50 \times 50$ screen (say), we then have that $\tau \approx 10^{-3} n$. We consider two cases, one with $n=25$ ($\tau \approx 0.03$), henceforth the \emph{low optical depth} case, and $n=250$ ($\tau \approx 0.3$), which we call the \emph{high optical depth} case. On a practical level, these two individual distributions are constructed by plucking points from bivariate Gaussians with relatively large variances over a disc of diameter $50 R_{E}$. For each of these two cases, we used the available sampling functions in \small MATHEMATICA\textregistered \, \normalsize to define the macrolens. %Note that for a collection of solar mass and solar radii lenses, one can estimate that $U \lesssim G M_{\odot} n /c^2 R_{\odot} = 2 \times 10^{-6} n$, and therefore the weak-lensing assumptions given in Sec. 3 are comfortably satisfied for $n \ll 10^{6}$. 
The resulting initialisations are shown in Figures \ref{fig:grav25} and \ref{fig:grav250}, respectively. In these Figures, the overlaid colour scale shows the projected potential $\psi(\boldsymbol{x})$ from expression \eqref{eq:psipot}, which is naturally stronger by a factor $\sim 10$ in the higher $\tau$ case. 

It should be stressed that the distributions considered here are not meant to represent any particular astrophysical system. They are constructed primarily to illustrate the mathematical machinery and to qualitatively explore what the wave-optical impact of lensing by $n \gg 1$ stars may be for continuous GWs in the $\sim$kHz band. %which we anticipate may be detected in the near future.

\section{Picard-Lefschetz approach}

As mentioned in the introduction, evaluating expression \eqref{eq:diffint} is challenging because the integrand oscillates an infinite number of times over the aperture (real plane). Standard numerical methods that involve finite cutoffs, for example, fail to return an adequate evaluation because, depending on whether one truncates at a trough or a crest of the time-delay $t_{d}$, the integral will either be under- or over-estimated, respectively. To make progress, we make use of the ideas behind PL theory, as described by \cite{feld19} \cite[see also][for alternative approaches]{diego19,guo20}.

We begin by introducing polar coordinates, $x = r \cos \theta$ and $y = r \sin \theta$, so that we have only one (semi)-infinite interval ($0 \leq r < \infty$) to consider. The transformed integral is evaluated for fixed values of $\theta$; given a set of values for $F(\boldsymbol{x}_{s},\theta)$, we can eventually build-up the full 2D integral using Simpson's (or some other standard) method because the angular limits are finite. The PL strategy begins by extending $r$ into the complex plane (i.e., analytically continuing the variable), viz. $r \rightarrow \text{Re}(\boldsymbol{r}) + i \text{ Im}(\boldsymbol{r})$, effectively doubling the number of (real) coordinates. Such an extension allows us to then deform the original integral into the complex plane. In particular, provided that the exponent $t_{d}$ is analytic (cf. Sec. 4.1), integrals around closed, complex contours will vanish by Cauchy's theorem, i.e.,
 \begin{equation} \label{eq:cauchy}
 \oint_{\Gamma} e^{2 \pi i \fGW t_{d}(\boldsymbol{r},\theta,\boldsymbol{x}_{s})} d \boldsymbol{r} = 0,
\end{equation}  
around any closed loop $\Gamma \subset \mathbb{C}$ (again, for a fixed $\theta$). If we were to thus design a contour consisting of multiple segments $\gamma_{i}$ such that $\Gamma = \sum_{i} \gamma_{i}$, the first of which ($\gamma_{1}$) is the non-negative real line, we can effectively evaluate the original integral \eqref{eq:diffint} by summing the remaining integrals over $\gamma_{i \geq 2}$.

The key observation now is that there is freedom in choosing these contours. Noting that the main issue with evaluating the original integral is its oscillatory nature, we proceed by choosing the contours precisely such that the oscillations are damped out as much as possible, so that standard numerical methods may be applied. 

In general, the function $t_{d}$ can itself be expanded into real and imaginary components, conventionally written as $i t_{d}(\boldsymbol{r},\theta,\boldsymbol{x}_{s}) = h(\boldsymbol{r},\theta,\boldsymbol{x}_{s}) + i H(\boldsymbol{r},\theta,\boldsymbol{x}_{s})$. Starting from some finite radius, we trace a \emph{Morse flow} $\boldsymbol{\gamma}(\lambda) = \{ \text{Re} [\boldsymbol{r}(\lambda)], \text{Im} [\boldsymbol{r}(\lambda)] \}$ according to \citep{witten11},
\begin{equation} \label{eq:morseflow}
\frac {d \gamma^{i}} {d \lambda} = - G^{ij} \frac {\partial h} {\partial \gamma^{j}},
\end{equation}
where $G^{ij}$ is a metric on the complex plane\footnote{Throughout this paper, we consider the Kronecker delta, $G^{ij} = \delta^{ij}$, by topologically associating $\mathbb{C}$ with $\mathbb{R}^{2}$. In some cases it may be advantageous to consider different metrics \citep{witten11}, but we ignore such generalisations here for simplicity.} and we have introduced an affine parameter $\lambda$ which labels the position along the curve. Along this particular contour, we have that
\begin{equation}
\begin{aligned}
\frac {d H} {d \lambda} &= \frac {d \gamma^{i}} {d \lambda} \frac {\partial H} {\partial \gamma^{i}} \\
&= -G^{ij} \frac {\partial h} {\partial \gamma^{j}} \frac {\partial H} {\partial \gamma^{i}} \\
&= 0,
\end{aligned}
\end{equation}
where the last equality holds because of the Cauchy-Riemann equations. In effect, the above result demonstrates that the oscillatory portion of the integrand is constant along a Morse flow, and can thus be pulled outside of the integral. Furthermore, the Morse flow is also a contour of steepest descent in the sense that $h$ is always decreasing, $d h / d \lambda = - \sum_{j} (\partial h / \partial \gamma^{j})^{2}$. This is the power of the PL approach: not only are oscillations neutralised, the real portion also decreases as quickly as possibly and truncation can be exacted at low values of $\text{Re}(\boldsymbol{r})$ without sacrificing accuracy \citep{witten11,feld19}.

%MENTION ESTIMATE LEMMA.

In general, however, there will not be a single Morse flow over the entire domain, but rather it is necessary to consider a sequence of such flows. The reason for this is that if a point of stationary phase is encountered, the right-hand side of equation \eqref{eq:morseflow} vanishes, implying that the flow halts as the velocity $d \boldsymbol{\gamma} / d \lambda$ tends to zero. It is therefore necessary to attach a flow to each individual image, where the beginning (i.e., initial condition) of each segment corresponds to the end-point of the previous plus a small perturbation. Each such segment is referred to as a \emph{Lefschetz thimble} \citep{feld19,jow20,jow21}, the overall sum of which defines the contour we wish to integrate along; for mathematical details concerning the well-posedness of such flows, we refer the reader to \cite{witten11}.

\begin{figure}
  \includegraphics[width=0.49\textwidth]{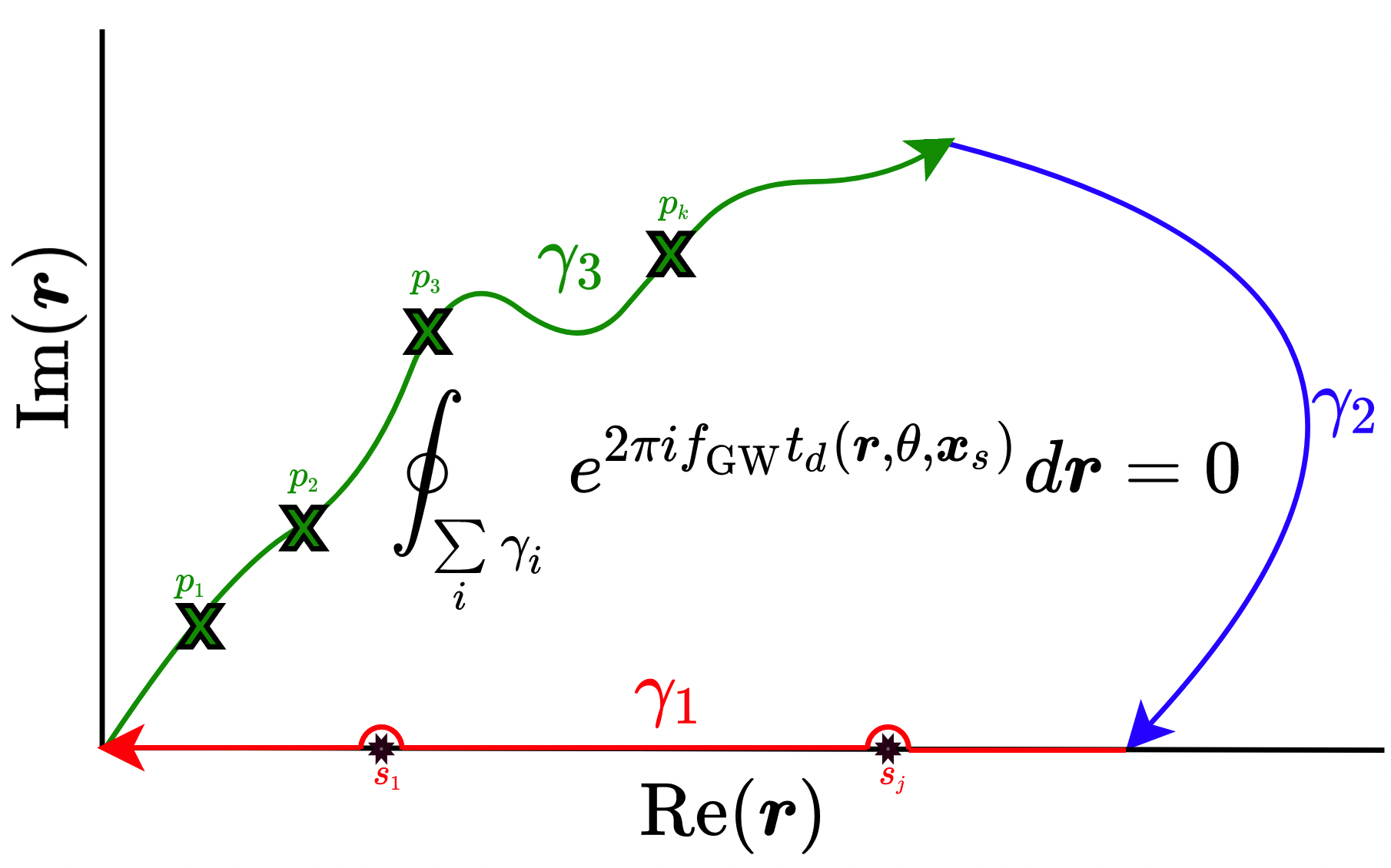}
  \caption{Graphical representation of Cauchy's theorem \eqref{eq:cauchy} as applied in the evaluation of the Fresnel-Kirchhoff diffraction integral \eqref{eq:diffint}. The integral of interest $(0 \leq \text{Re}(\boldsymbol{r}) < \infty)$ lies along (the negative of) $\gamma_{1}$, shown in red, though two additional lines are introduced to form a closed contour. {If the original segment contains branch points $s_{1}, \ldots, s_{j}$, these are exorcised from the domain to preserve analyticity (see Sec. 4.1).} A \emph{Lefschetz thimble} is built by flowing in the direction defined by the Morse equation \eqref{eq:morseflow}, until a point of stationary phase (i.e., an image, labelled $p_{1}$) is reached. At this point, the velocity of the flow effectively tends to zero, and it is necessary to introduce a perturbation to continue the thimble. This process of flow and perturbation is continued until all images along the route, the last of which is $p_{k}$, are moved past. The overall sum of these flows defines the contour $\gamma_{3}$ (green), which is then connected back to the real line through an arc ($\gamma_{2}$; blue). This latter integral vanishes, in many cases of interest, by Jordan's lemma.
}
  \label{fig:cont}
\end{figure}

In summary, we design a 3-component contour $\Gamma$ by sequentially flowing from the origin according to \eqref{eq:morseflow} ($\gamma_{3}$), eventually joining to the real axis by introducing an arc $(\gamma_{2})$, which then connects back to the origin ($\gamma_{1}$), defining a closed contour. Furthermore, the integral along the arc $\gamma_{2}$ vanishes, under reasonable assumptions, by Jordan's lemma. Some additional numerical details are given in the next section, while a worked example pertaining to the generalised Fresnel integral is given in the Appendix A. Figure \ref{fig:cont} illustrates the core ideas involved in the PL evaluation. 

%The method described above can be used in the evaluation of the Fresnel-Kirchhoff integral \eqref{eq:diffint}.
%The flow continues until a second image ($p_{2}$) is met.

\subsection{Numerical implementation}

In practice, there are several numerical obstacles encountered when applying the above ideas. The first major difficulty concerns the fact that the Morse flow terminates at points of stationary phase. After performing the complex decomposition of the integrand, we locate the roots, parameterised by the angle $\theta$ and the screen parameters $\boldsymbol{x}_{s}$, of the first derivative of $t_{d}$ plus the logarithm of the Jacobian\footnote{{Such a step is not strictly necessary, as one could instead monitor the flow to check if an image has been met (i.e., if the velocity falls below a threshold), and automatically issue a perturbation in the manner described to continue the thimble. For the $n \lesssim 250$ cases considered here, this root solving stage is relatively inexpensive, though for $n \gg 10^2$ a subroutine approach would be faster.}}. This entails solving high-order polynomial equations, which we achieve through an exhaustive search using different initial guesses and the Newton-Raphson method within \small MATHEMATICA\textregistered \normalsize. Once an image is encountered, the flow velocity is then artificially perturbed, $\boldsymbol{\gamma}'(\lambda) \rightarrow \boldsymbol{\gamma}'(\lambda) + \epsilon$, to `kick' the flow onto the next thimble. In practice, we set $\epsilon = 10^{-4}$. 

Not all images are of relevance, however, since some may be topologically disconnected from the overall flow (see Appendix A). For example, if a saddle occurs at a place of negative real component $[\text{Re}(\boldsymbol{p}) < 0]$, the flow cannot encounter it and the associated thimble is irrelevant. Classifying such irrelevant images and related Stokes transitions is in general a difficult problem; see \cite{feld19} for a thorough discussion. We employ something of a trial-and-error approach in this paper, where images are flowed from, but only connected (relative to the origin) components are kept to ensure topological continuity. 

% This value was chosen because the result, tested at several values of $\boldsymbol{x}_{s}$, differed negligibly from the value when using 255 or 257 steps.

{The root solver is initialised over a grid of $\theta$ and $\boldsymbol{x}_{s}$ values, and a set of thimbles (in $\boldsymbol{r}$) is then constructed at each grid point. For each figure produced here, $512$ angular steps are used, i.e., a spacing of $2 \pi / \Delta \theta = 2 \pi / 512$ is used. The results obtained with this resolution differ by at most $\sim 1\%$ from those obtained with $256$ steps. Some additional tests on convergence with respect to resolution are given in Appendix B. Generally, higher radiation frequencies require larger $\Delta \theta$ (by a factor $\sim f / \fGW$) because the integrand varies more rapidly, and thus $\sim 128$ steps is already adequate for frequencies $\lesssim 1$kHz. Simpson's method is then used to sum the $\theta$-integrals together to build the full, 2D diffraction integral \eqref{eq:diffint}. }

The Morse flow equations \eqref{eq:morseflow} are sequentially solved using a Runge-Kutta method up to some maximum radius $\text{Re}(\boldsymbol{r})$. Formally speaking, this radius must extend to infinity, else Cauchy's theorem \eqref{eq:cauchy} cannot be applied. However, because the Morse flow is also a contour of steepest descent, high accuracy can be achieved even with relatively early truncations that depend on $\boldsymbol{x}_{s}$; see Appendix A. %The end point is computed iteratively in this work, where the flow is terminated once the integral no longer increases by a small threshold value.

%NEED TO CORRECTLY MENTION THAT BRANCH POINTS LIE ON THE REAL AXIS!

Finally, it is important to note that the function $t_{d}$ is not analytic over the complex plane, as the $\psi$ piece introduces logarithmic singularities at each microlens position $(r_{k} \cos \theta_{k}, r_{k} \sin \theta_{k})$. This is a problem {firstly because Cauchy's theorem cannot be strictly applied for $\theta=\theta_{k}$, though small arcs around the branch points $r_{k}$ can be constructed to restore analyticity if such an angle is within the numerical grid.} Additionally, the flow tends to infinitely wind around singular points (i.e., the velocity blows up), causing the integral to diverge. We have explored several possibilities for tempering the singularities:
\begin{itemize}
\item{Introducing an additional $n-1$ lens planes can circumvent the appearance of more than one singularity on any given plane, as discussed by \cite{feld20b} \cite[see also][]{ram21}. In particular, if each microlens resides on a different plane, screen-adapted coordinates can be used to center each individual singularity away from the relevant regions, so that it may be ignored. This approach introduces considerable computational demand however, as the Fresnel-Kirchhoff integral effectively becomes $2n$ dimensional.}
\item{Various regularisation techniques are possible, including (i) approximating $\psi$ near the singularities by a different function that is regular there \cite[cf.][]{christ18,guo20}, and (ii) cutting out regions of some size surrounding the singularities. These approaches can be difficult to tune however because if too much of the contour is cut, or if $\psi$ is poorly approximated, significant errors can be introduced.}
\item{In the building of the total contour $\gamma_{3}$ and applying \eqref{eq:cauchy}, it is not necessary that the entire portion (or even any portion) be a Morse flow. Therefore, if singularities lie within the Morse-constructed contour, one can deform the path to restore analyticity, as is typically done when computing inverse Laplace transforms, for example. These deformed segments could be fittingly called `suboptimal' thimbles.}
\end{itemize}

A thorough exploration of the above possibilities lies beyond the scope of this paper. Nevertheless, the third option is employed here for concreteness, as some testing with low $n$ cases suggests the results agree with the multi-plane approach to within a few percent. Since it is only ever necessary to suboptimally flow over short segments, the oscillations introduced are not fatal for the numerical method. %The simplest such suboptimal thimble, which we use here, is just a horizontal line of constant imaginary coordinate. %RE

%For cases with singularities near the origin, we can similarly begin the flow at some finite radius away from the origin, 

%The simplest such suboptimal thimble is of course a straight line, and for cases with singularities near the origin it may be simpler to begin the flow at some finite $\text{Re}(\boldsymbol{r})$.

\section{Results}

Having introduced the PL approach, we are now in a position to evaluate expression \eqref{eq:diffint} for the microlens distributions described in Sec. 3.2, i.e., for a sparse cluster (Sec. 5.1) and a dense cluster (Sec. 5.2).

\subsection{Low optical depth}

\begin{figure}[h]
\centering
  \includegraphics[width=0.49\textwidth]{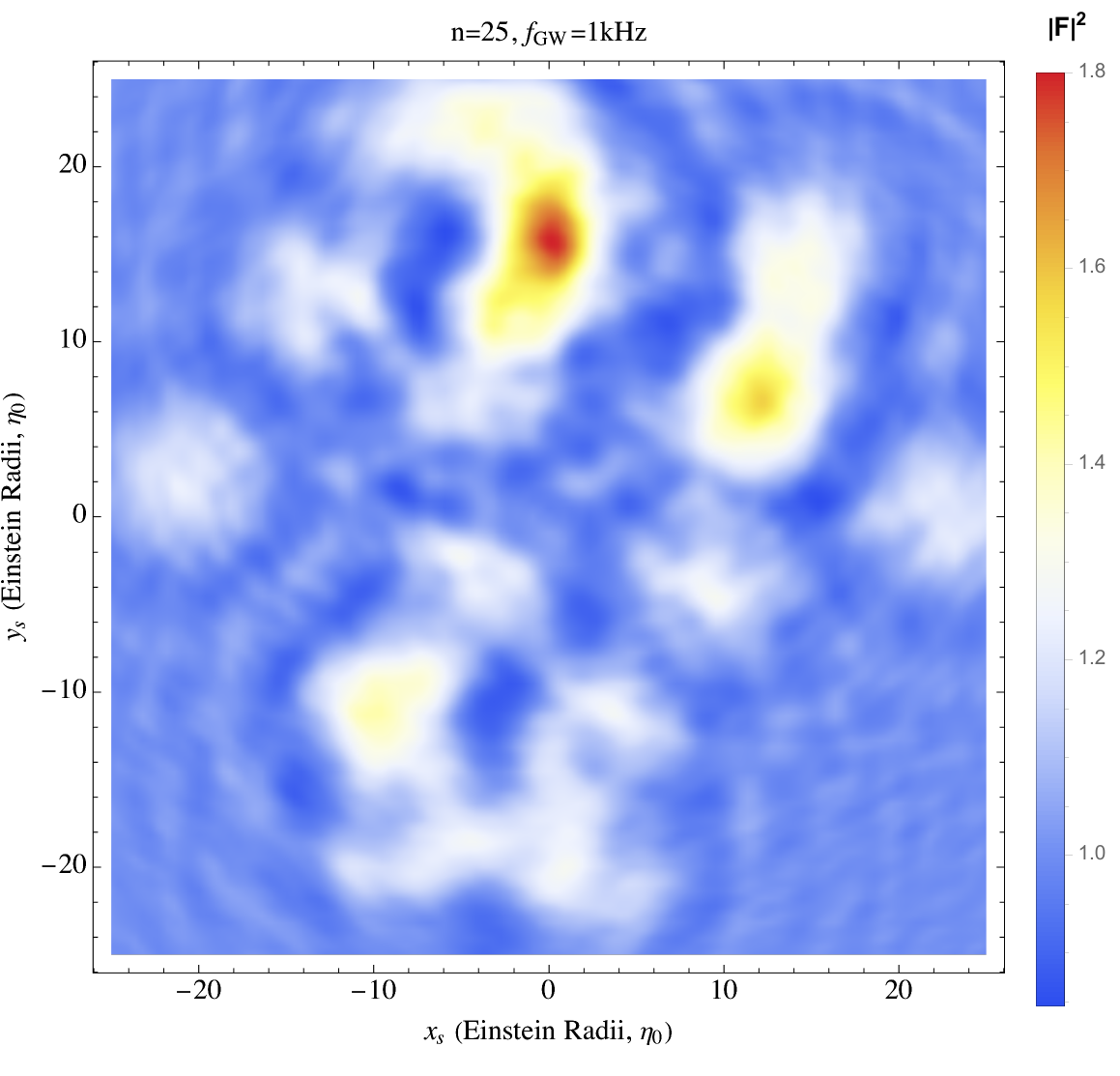}
  \caption{The intensity profile, $|F(\boldsymbol{x}_{s})|^2$, associated with the gravitational potential shown in Fig. \ref{fig:grav25} for $\fGW = 1$ kHz. Hotter shades indicate greater amplifications. {The resolution is $110 \times 110$ (cells).}
}
  \label{fig:intensity2d25}
\end{figure}

\begin{figure}[h]
\centering
  \includegraphics[width=0.49\textwidth]{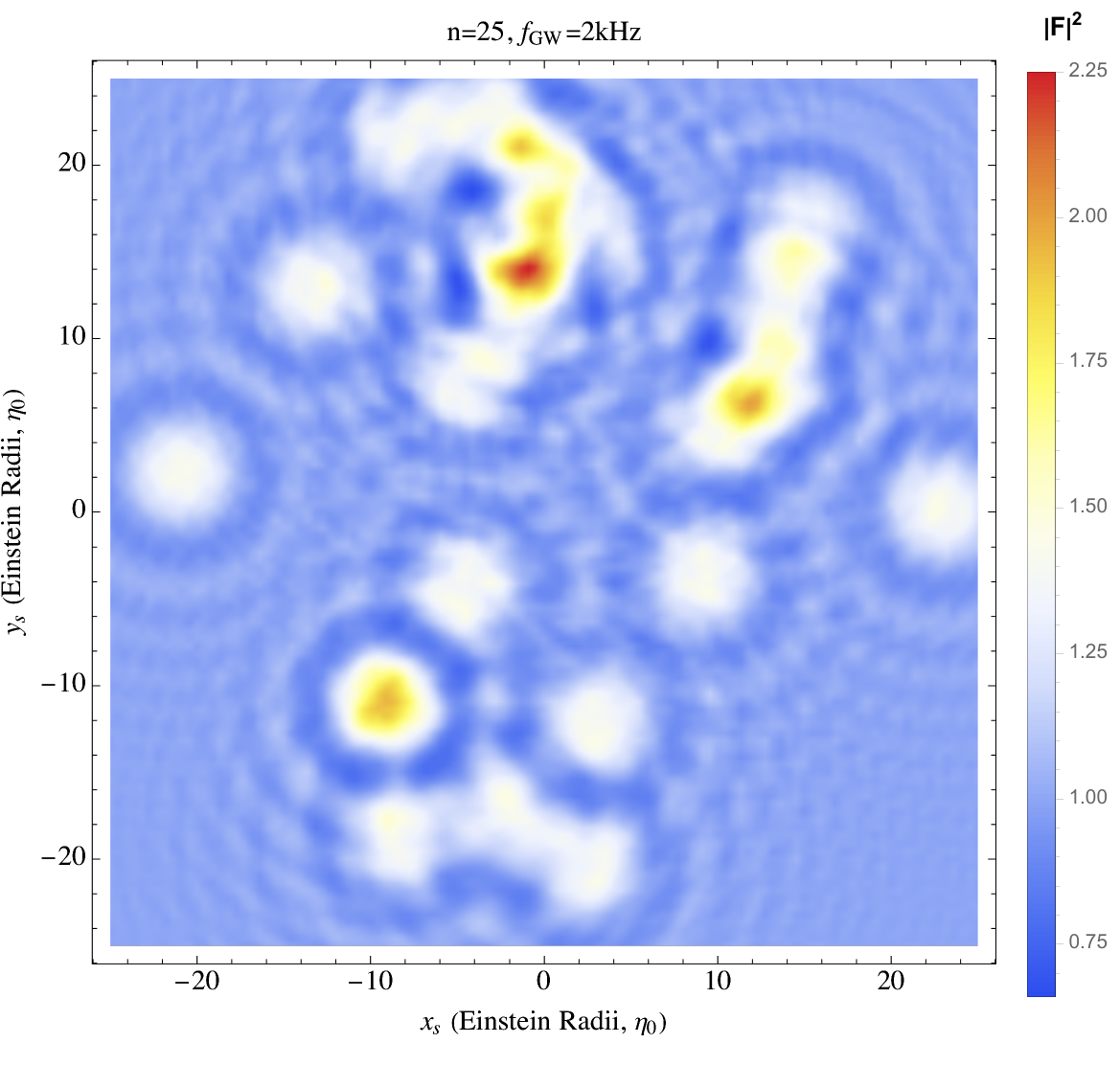}
  \caption{Similar to Fig. \ref{fig:intensity2d25} though for $\fGW = 2$ kHz.
}
  \label{fig:intensity2d252}
\end{figure}

Figure \ref{fig:intensity2d25} shows the intensity pattern, as a function of the normalised screen parameters $x_{s}$ and $y_{s}$, associated with the low-optical depth microlens distribution shown in Fig. \ref{fig:grav25}, where we have fixed $\fGW = 1$ kHz. Figure \ref{fig:intensity2d252} instead shows the same intensity profile but for $\fGW = 2$ kHz, i.e., for a star spinning twice as quickly. Although no neutron stars with such a high rotational velocity have been directly observed, this latter case provides a useful comparison to illustrate how the radiation frequency impacts on the overall intensity map. Furthermore, theoretical models of accretion-induced spin-up allow, in principle, for the rotation rate to reach this level \cite[e.g.,][]{glamp21}, and GW searches at this frequency range have been conducted \citep{derg21}.

For $\fGW = 1$ kHz, the maximum intensity over the screen is relatively low, $|F|_{\text{max}}^2 = 1.8$, owing primarily to the sparse nature of the microlens distribution and, hence, the weakness of the bulk gravitational potential $U$. For the faster star with $\fGW = 2$ kHz, we instead have $|F|_{\text{max}}^2 = 2.3$. Monotonicity in $|F|_{\text{max}}$ as a function of frequency is generally expected, since as $\fGW \rightarrow \infty$ we approach the geometric optics limit, where the amplification becomes formally infinite along the caustic surface(s) where $\boldsymbol{x} + \tfrac{1}{2} \nabla_{\boldsymbol{x}} \psi = 0$ \citep{jow21}. To put these maxima in perspective, consider the case of a single solar-mass lens, where one finds maximum values of $|F_{n=1}|^2 = 1.21$ and $|F_{n=1}|^2 = 1.44$ for $\fGW = 1$ kHz and $\fGW = 2$ kHz, respectively {[see equation \eqref{eq:formula}]}. These maxima occur when the source is oriented directly behind the perturbing body, i.e., at $(x_{s},y_{s}) = (x_{1},y_{1})$. It is unsurprising therefore that the maxima in our simulations occur when the source aligns itself behind the centre of mass of the densest mini-cluster located at $(x_{s},y_{s}) \approx (0,15)$; see Fig. \ref{fig:grav25}. In particular, a collection of point masses in close proximity to one another generally behave as one larger lens around the centre of mass, and since $|F|$ scales with $M_{L}$, as can be seen from \eqref{eq:diffint}, the amplification is larger there. In the local vicinity (i.e., within a few Einstein radii) of isolated stars in our distribution, such as at $(x_{s},y_{s}) \approx (-20,0)$, the intensity strongly resembles that of the single point-lens case \citep{liao19,meena20}.

The oscillatory nature of the intensity along any given line is also the hallmark of interference, as expected in a wave-optics calculation. The variability is more extreme in the higher frequency case in Fig. \ref{fig:intensity2d252}, as the dimensionless exponent $\fGW t_{d}$ varies over length-scales exactly half as long. The emergence of interference fringes is also more obvious in this case, especially surrounding isolated members of the cluster, e.g., near $(x_{s},y_{s}) \approx (-20,0)$. The bulk influence is minimal in the vicinity of these regions, and the amplification roughly matches the analytic profile for the single lens case described above, as does the spacing between interference fringes computable from the Fourier spectrum \citep{nak99}. For lower GW frequencies, the amplitudes of the oscillations become virtually invisible since the wavelength is so long that the GWs hardly experience the lens, implying that lensing by stellar-mass bodies is unimportant for `ordinary' neutron stars with $\nus \lesssim 10^2$ Hz.

\begin{figure}[h]
\centering
  \includegraphics[width=0.49\textwidth]{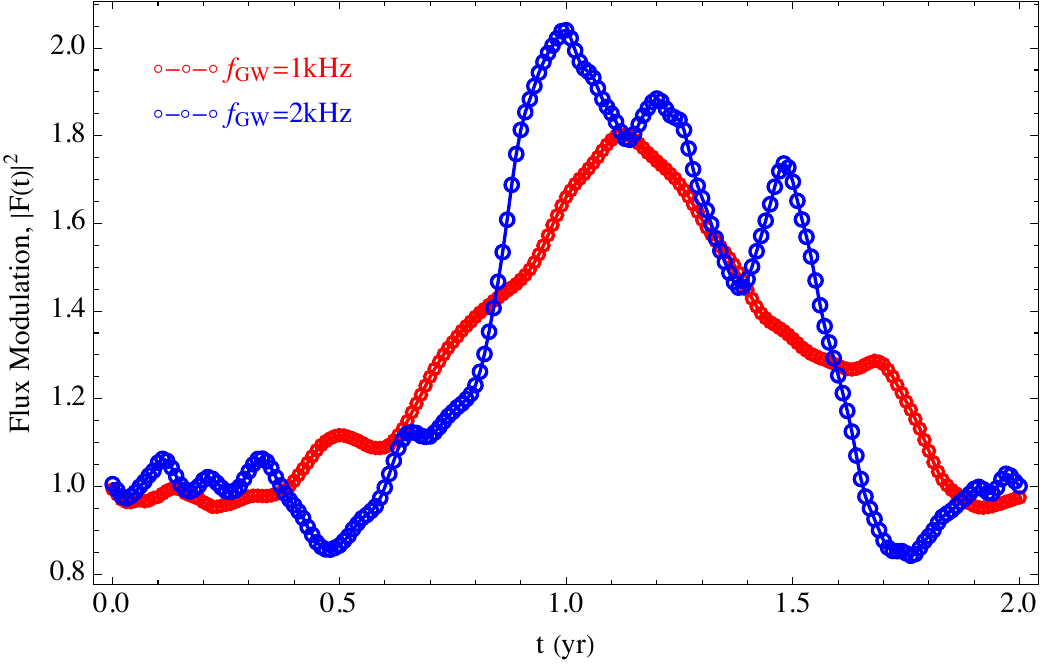}
  \caption{The flux modulation, $|F(t)|^2$, observed in the case of a source moving in the $+y_{s}$ direction relative to the origin at $t=0$ for the microlens distribution shown in Fig. \ref{fig:grav25}. The red (blue) symbols correspond to $\fGW = 1 (2)$ kHz, with their size roughly representing the maximum level of numerical error present in the calculation. The speed of the neutron star is taken to be $v = 600 \text{ km s}^{-1}$, so that it crosses $\sim 14 \eta_{0}$ per year.
}
  \label{fig:intensity25}
\end{figure}

\begin{figure}[h]
\centering
  \includegraphics[width=0.49\textwidth]{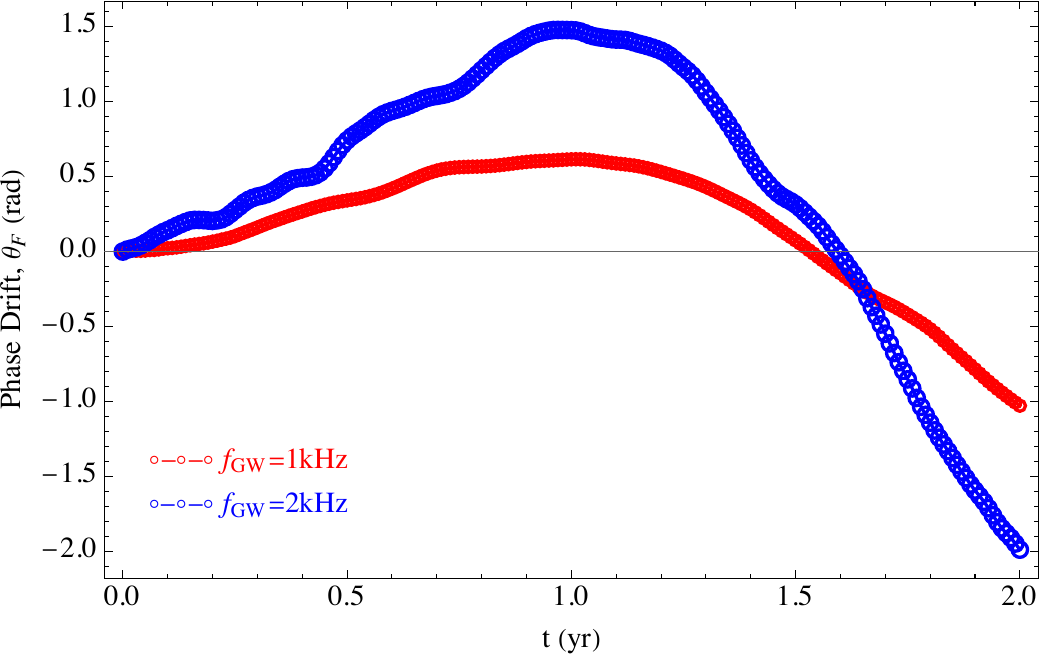}
  \caption{Similar to Fig. \ref{fig:intensity25}, though instead showing the phase drift $\theta_{F} = - i \ln (F/|F|)$, offset such that $\theta_{F}(t=0)=0$.
}
  \label{fig:phase25}
\end{figure}

To better illustrate the wave-like nature of the Fresnel-Kirchhoff integral, we consider also flux, $|F(t)|^2$, and phase, $\theta_{F}(t) = - i \ln [F(t)/|F(t)|]$, variations along a hypothetical trajectory of the source. To this end, suppose that the neutron star begins at the origin $(x_{s},y_{s}) = (0,0)$ at $t=0$, and then moves, relative to the macrolens, in the $+y_{s}$ direction with velocity $v = 600 \text{ km s}^{-1}$; a value not unreasonable for millisecond pulsars \citep{hobbs06}. (Note, however, that a smaller $v$ but longer $T_{\text{obs}}$ or $D_{\text{OL}}$ yields the same qualitative picture). Within a year, the star therefore travels $\sim 14 \eta_{0}$ [see equation \eqref{eq:units}], crossing a large number of interference fringes. Figures \ref{fig:intensity25} and \ref{fig:phase25} show the 1D variations in flux and phase, respectively, experienced along this path as a function of time, where we implicitly ignore variations in $D_{\text{OS}}$. The red (blue) data points correspond to $\fGW = 1 (2) \text{ kHz}$, with the size of the symbols roughly characterising the error bars present in the calculation ($\lesssim$ percent level).

The oscillatory nature of the modulation is evident in the flux, especially near $t=0$ where the source is caught between two neighbouring elements of the cluster (the origin in Fig. \ref{fig:grav25}). In regions of low density, i.e., at early or late times, the $\fGW =2$ kHz case oscillates roughly twice as often, as expected from the relative decrease in spacing between interference fringes. The maximum flux achieved by the higher-frequency case is greater than in the lower frequency case by $\sim 25\%$, in accord with Figs. \ref{fig:intensity2d25} and \ref{fig:intensity2d252}. These maxima are achieved after approximately one year of travel time in this setup (or two years with $v \sim 300 \text{ km s}^{-1}$), whereupon the neutron star enters into the densest region shown in the top half of Fig. \ref{fig:grav25}. % These figures demonstrate that, depending on where the neutron star is located, non-negligible amplitude (factor $\lesssim 2$) and phase variations could be expected in continuous GW signals over sufficiently long observational runs.

Similar oscillatory patterns are observed in the phase, which covers a wide range of values ($-2 \lesssim \theta_{F} \lesssim 1.5$) over $T_{\text{obs}}$. Because phase drifts exceeding a sizeable fraction of unity are likely to inhibit a fully coherent search \citep{jones04,dre18}, variations on the order seen in Fig. \ref{fig:phase25} are potentially detrimental for detection prospects, despite the flux enhancement (Fig. \ref{fig:intensity25}). Within $\sim$ 6 (3) months, the phase wanders by more than $0.3$ radian for the $\fGW = 1 (2)$ kHz source in this scenario, though the gradient $d \theta_{F} / d t$ increases more rapidly in the dense parts of the cluster and coherence is lost at a faster rate. As such, it is likely that the data would need to be analysed semi-coherently in this scenario over (at most) $\sim$~month-long segments, especially in the higher frequency case, even in the absence of (accretion-induced or otherwise) spin wandering. A thorough examination of the trade-off between lensing-induced amplification and decoherence lies beyond the scope of this work; we refer the reader to \cite{suva16,dre18} for a comparison between fully- and semi-coherent sensitivities. Given a lens model however, such phase errors may be corrected for in the template waveform using the PL methodology described here. Either way, issues related to phase modulation may be further compounded by the Earth's diurnal and orbital motions if the sky location or the relative velocity of the source is poorly constrained. 

%citep{suva16}. There is a therefore a trade-off between the enhanced flux (Fig. \ref{fig:intensity25}) imparted by lensing and the loss of coherence due to phase variations

%and at $(x_{s},y_{s}) \approx (0,5)$ where there is instead a break in the interference ring

\subsection{High optical depth}

\begin{figure}[h]
\centering
  \includegraphics[width=0.49\textwidth]{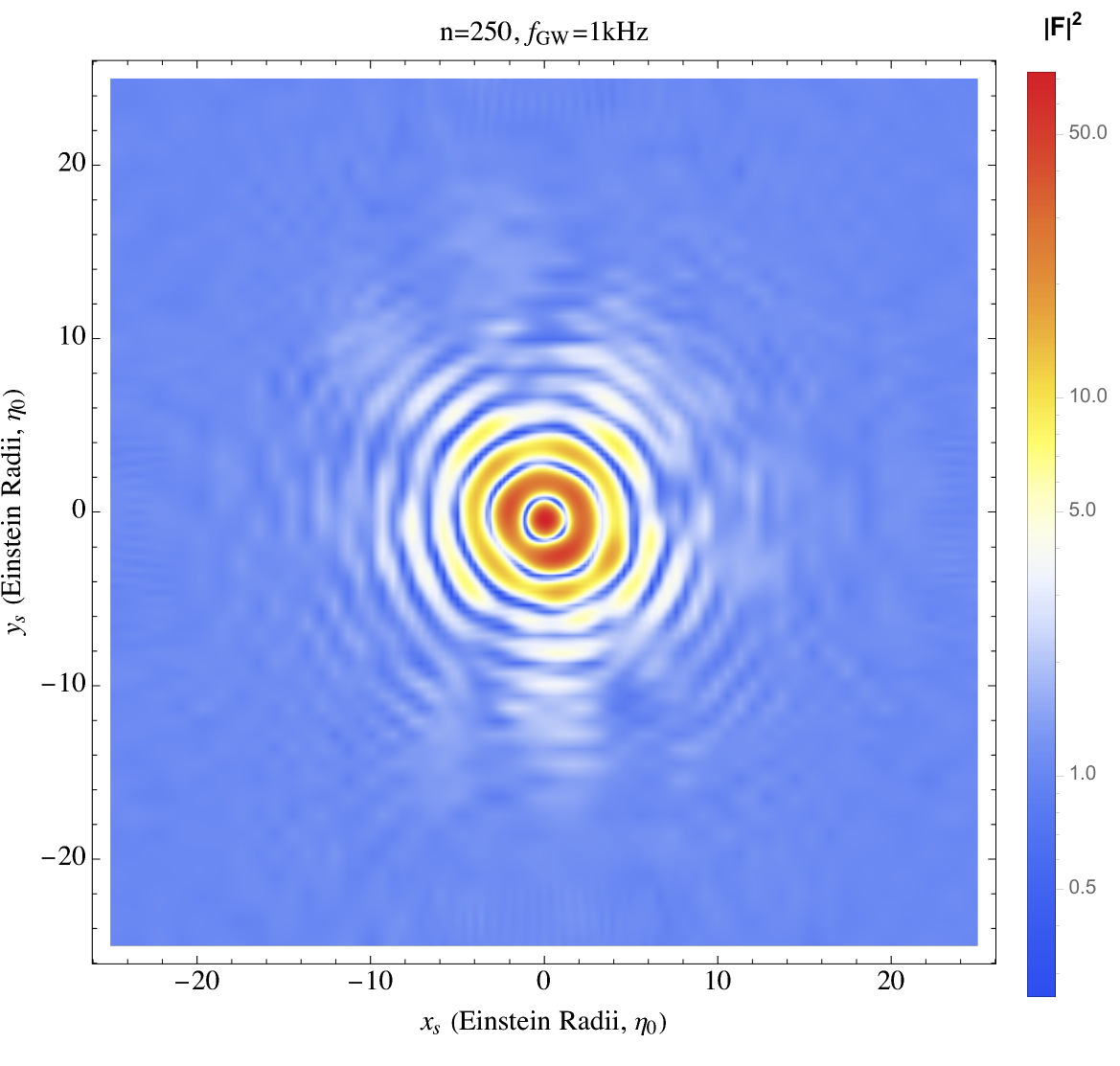}
  \caption{Similar to Fig. \ref{fig:intensity2d25}, though for the high optical depth case shown in Fig. \ref{fig:grav250}. {The (logarithmic) resolution is $100 \times 100$ (cells).}
}
  \label{fig:intensity2d250}
\end{figure}

Similar to the previous section, Figure \ref{fig:intensity2d250} shows the intensity pattern with $\fGW = 1$ kHz, as a function of the screen parameters, for the $n=250$ case shown in Fig. \ref{fig:grav250}. Because $\psi$ is everywhere larger by a factor $\sim 10$ than in the previous case, and more importantly because the distribution is highly concentrated around the origin, the bulk magnifications in the heart of the cluster are considerably larger ($|F|_{\text{max}}^2 = 73$). By contrast, the analytic formula {(see Appendix B)} for a point-mass lens with $M_{L} = 250 M_{\odot}$ gives $|F_{n=1}|^2 = 97$ at the origin for the same radiation frequency. For $n \gtrsim 10^2$ and the Gaussian distributions we use, the overall mass density on the screen appears similar to that of a grainy sphere with $\psi$ decaying with radius as a power-law. Over- and under-dense regions in the magnification profile are clearly visible however even for $n=250$, especially at $(x_{s},y_{s}) \approx (2,-2)$ where there is angle-dependent structure within the second peak `ring'. Interference fringes are also abundant, as can be seen from the network of graininess near the origin that extends out to $|\boldsymbol{x}_{s}| \lesssim 20$. Overall, the intensity profile is compactified over the screen relative to the $n=25$ case; in the limit $n \rightarrow \infty$, $\psi$ tends to that of a point-lens with ever-growing $M_{L}$ \citep{nak99}.

\begin{figure}[h]
\centering
  \includegraphics[width=0.49\textwidth]{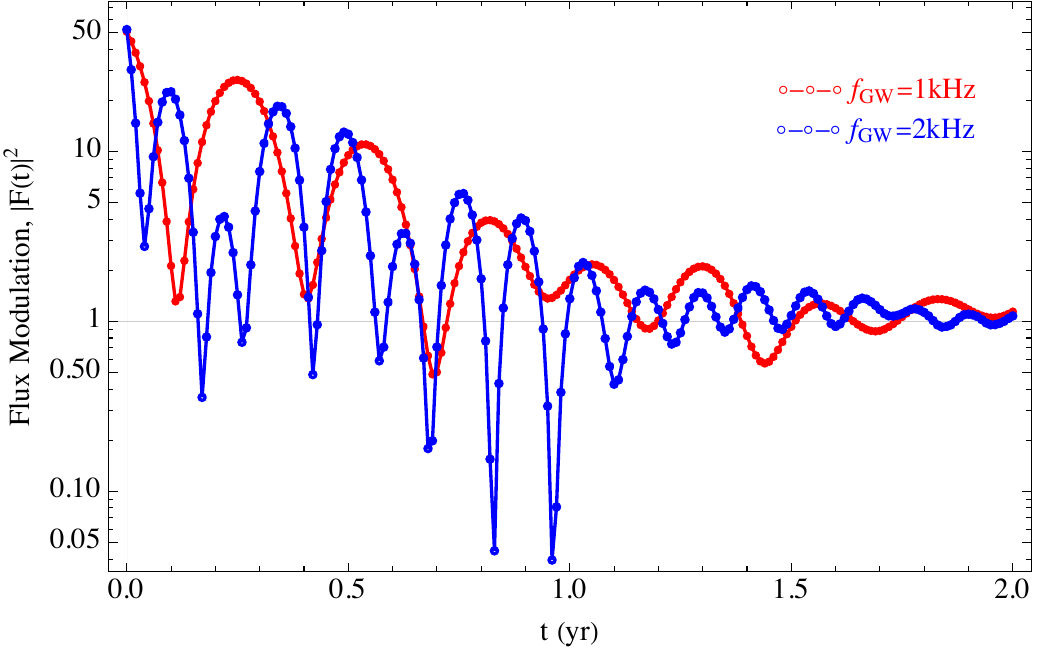}
  \caption{Similar to Fig. \ref{fig:intensity25}, but for the high optical depth distribution shown in Fig. \ref{fig:grav250}. The source velocity is taken to be $v = 300 \text{ km s}^{-1}$ in the $+y_{s}$ direction. The numerical errors roughly match those of the $n=25$ simulation ($\lesssim$ percent level), though appear smaller due to the logarithmic scaling of the vertical axis.
}
  \label{fig:intensity250}
\end{figure}

%In the lower frequency case, we see that, as the star moves out of the densest region of the cluster, the flux rises almost monotonically, eventually decaying at the same rate it rose after $\approx X$ yr.

To better visualise the oscillations above the ambient contribution, suppose that the source moves in the $+y_{s}$ direction with velocity $v = 300 \text{ km s}^{-1}$ (or greater $v$ and proportionately contracted time axis). Similar to Fig. \ref{fig:intensity25}, the variation in flux for the $n=250$ cluster is shown in Figure \ref{fig:intensity250}, where the red (blue) symbols correspond to $\fGW = 1 (2)$ kHz. 
The emergence of interference fringes is readily apparent as the flux begins at a maximum as the line of sight intersects with the densest region at the origin, descends to an order-unity trough after $\lesssim 2 (1)$ months in the $\fGW = 1 (2)$ kHz case, and then rises on a similar timescale to reach a second peak. Each subsequent peak after the first, the total number of which is doubled in the higher frequency case as expected, is generally of lower amplitude because the stellar density decreases as a function of radius (see Fig. \ref{fig:grav250}). This pattern is not monotonic however, especially around $t \sim 4$ months for the $2$~kHz case where only a mild peak is reached ($|F|^2 = 4$), because the lens distribution is grainy. In this case, the phase drifts are so extreme that coherence is likely to be lost even over timescales of $\sim$~weeks. Once the line of sight encounters only the outskirts of the cluster after $\sim 1.5$ yr, the flux closely resembles that seen in Fig. \ref{fig:intensity25}, though oscillates more frequently as there are a greater number of interference fringes produced by the microlenses. %The overall amplification is marginally larger in the higher frequency case. %For the extreme cases shown here in this Section, the phase drifts by more than 1 radian even on sub-month timescales, and 

%This latter case, while still within the diffractive regime, exhibits interference effects as the intensity curve is clearly non-monotonic over even $\sim$month-long time-scales.

\subsection{Implications for parameter estimation}

%especially given the absence of detections thus far.
%depending on how well the oscillations of $|F|$ can be resolved in the detector

Unlike in the case of a merging binary, where the bulk of the GW luminosity is produced within a few seconds, continuous GW emissions from a deformed neutron star are longer lived -- potentially persisting for its entire lifetime -- though much fainter; see equation \eqref{eq:massquadamp}. With present-day (future) instruments, it is likely therefore that a detection of these sources requires monitoring for a year (month) or more \citep{suv21,sold21}. If the line of sight linking the neutron star to the detector intersects with a dense cluster during the observation, lensing effects may modulate the strain, as explored in Figs. \ref{fig:intensity25} and \ref{fig:intensity250}. What impact could this be expected to have on parameter estimation? 

% thereby measuring the $\boldsymbol{x}_{k}$ and $M_{k}$ parameters,
For magnetic deformations, the gravitational waveform is expected to have a sinusoidal profile of the form $h(t) \sim h_{0} \sin (2 \nus t)$ plus some subdominant harmonics. The lensed waveform will be similar but with a {time-dependent and complex prefactor, so that the signal is amplified by a real factor while its phase is shifted}. Because the variations in $h(t)$ are much more rapid than the interference-induced variations in $F(t)$, it may be possible to disentangle the effects of lensing via `beat' patterns \citep{jung19,meena20,cheung21}. At the simplest level however, we note that only mean values for the (upper-limit) quadrupole moments of neutron stars are typically reported from search efforts \citep{derg21,ligo21}. This is because the coherent nature of continuous GW-searches necessitates that averaging processes between individual interferometers be carried out to properly subtract noise \citep{jaran98,owen10}. As such, a detection of the mean strain $\langle h_{0} \rangle$ implies that the true strain would be lower by a factor $1/\langle |F| \rangle$, where the angled brackets denote a statistical average over $T_{\text{obs}}$. {For small clusters}, the amplification will not exceed $\sim 20\%$ or so (see Fig. \ref{fig:intensity25}), implying that, since $h_{0} \propto B_{\star}^2$, the difference in the $B$-field estimate would likely be no more than $\sim 10\%$. {For $n\gg25$, the amplifications could potentially be much larger \cite[see Fig. \ref{fig:intensity250} and][]{mishra21}.}  %Still, this may couple

The signal-to-noise ratio (SNR) of the system, given by \citep{jaran98}
\begin{equation} \label{eq:snr}
\text{SNR} \simeq \sqrt{ \frac{2}{S_{n}(\fGW)} \int^{T_{\text{obs}}}_{0} [h(t)^2] dt },
\end{equation}
may non-negligibly increase for large amplifications. Considering the trajectory defined within Fig. \ref{fig:intensity25}, for example, we estimate that the relative increase in the SNR due to lensing for $\fGW = 1$ kHz and $T_{\text{obs}} = 2$ yr reads
\begin{equation} \label{eq:snrex}
\begin{aligned}
\frac {\text{SNR}(F \neq 1)} {\text{SNR}(F = 1)} &\approx \sqrt{\frac { \int^{T_{\text{obs}}}_{0} |F(t)|^2 h_{0}^2 \sin^2(\fGW t) dt} {\int^{T_{\text{obs}}}_{0} h_{0}^2 \sin^2( \fGW t) dt}} \\
&= 1.13.
\end{aligned} 
\end{equation}
For the first $T_{\text{obs}} = 1$ yr, we find instead $\text{SNR}(F \neq 1)/\text{SNR}(F = 1) = 1.07$. While these SNR increases are marginal, they could be sufficient to propel an otherwise borderline source into the detectable threshold \citep{lasky15}. As seen in Fig. \ref{fig:phase25} however, these increases may be offset by the fact that coherence is lost due to lensing-induced phase wandering. Depending on the gradient $d \theta_{F} / d t$, it may be necessary to break the data into many shorter segments, thereby actually reducing the overall sensitivity \citep{suva16,dre18}. {For greater values of the total lensing mass over a fixed area, we generally find that the SNR increase is higher.} %for the $1$ kHz case depicted in Fig. \ref{fig:intensity250}, for instance, we have $\text{SNR}(F \neq 1)/\text{SNR}(F = 1) = X$.

For large SNR ($\gtrsim 10$), parameter-estimation errors that may result if lensing were not taken into account can be calculated through a Fisher analysis \citep{tak03}. Given the inner product $( \cdot | \cdot )$, well approximated by the integral in expression \eqref{eq:snr} for a monochromatic source \citep{jaran98}, one can define the Fisher matrix, $\Gamma_{ij} = \left( \frac {\partial h} {\partial q^{i}} | \frac {\partial h} {\partial q^{j}} \right)$, where $\boldsymbol{q}$ is a vector of parameters, which includes the source parameters (e.g., $\varepsilon$), location parameters (e.g., position relative to solar system barycenter), and lensing parameters (e.g., $M_{k}$). In general, these are not all independent. For example, the intensity pattern depends intimately on both $\fGW$ and the $M_{k}$. Regardless, the relative error on any given $q^{k}$ is estimable through $(\Delta q^{k})^{2} = (\Gamma^{-1})_{kk}$. Unfortunately, a reliable calculation of $\Gamma_{ij}$ requires one to build many amplification profiles so that derivatives with respect to the lens parameters can be taken, which is beyond our current computational capacity. Nevertheless, using the methodology described herein, a thorough exploration of the parameter space and relative errors could be performed; see Appendix A in \cite{tak03} for a Fisher analysis in the case of a single point-mass lens. %For instance, 

Finally, we note that a misinterpretation of a time-varying $h_{0}$ as being intrinsic to the source, as opposed to being a result of lensing, may have important consequences for the evolutionary modelling of the system. For example, in actively accreting systems that nevertheless show little variation in the spin frequency, it has been argued that GW backreaction may be a key factor in maintaining spin equilibrium \citep{bild98,git19}. The validity of this scenario can be tested, for any given system, if the braking index, $n_{b} = \nus \ddot{\nu}_{\star} / \dot{\nu}_{\star}^2$, can be independently measured: GW-dominated spindown implies $n_{b} \approx 5$. However, if $\varepsilon$ were to intrinsically vary over sufficiently short timescales, $n_{b}$ could differ significantly from 5, even for GW-dominated emission \citep{ara16}. Simultaneous measurements of $n_{b}$ and $\varepsilon$ could then be used as an independent means to quantify the wave-like effects of lensing.

\subsection{Larger clusters}

{Although we have considered $n \leq 250$ here, the PL approach can, in principle, be used to study wave-optical microlensing for arbitrary $n$. While lensing by a large number of stars is unlikely for a Galactic source, the amplification can be substantially increased if the perturbers are embedded into larger mass distributions \citep{mishra21,cheung21}, the extent to which is quantified by the optical scalars $\kappa$ and $\gamma$. For extragalactic sources, these quantities may be sizeable fractions of unity \citep{lew10,lew20}. As noted in Sec. 5.2, and although dependent on the source kinematics, phase drifts exceeding $\sim 1$ rad can theoretically occur within the space of a few weeks for $n=250$. For even larger $n$, the loss of coherence will be quicker, which may present a significant impediment to narrow-band searches as the signal is scrambled. 
%It would also be worth including the optical scalars $\kappa$ and $\gamma$, which might be sizeable fractions of unity in realistic scenarios \citep{lew10,lew20}.
}

\section{Discussion}

It is hoped that continuous GW signals from (magnetically-deformed or otherwise) neutron stars may be detected in the near future. Owing to their comparatively weak strain \eqref{eq:massquadamp}, a measurement likely requires observational monitoring for at least a year with current technology \citep{suv21,sold21}, during which the relative motion between the source and the detector is important, especially if matched filtering is to be carried out \citep{jaran98,dre18}. Although rare, it is possible that intermittent lensing by one or more stars may take place for sources located within/behind the Galactic bulge or a globular cluster, modulating the resulting signal to a degree that depends on the GW frequency, the lens distribution, and the source velocity \citep{pac86b,dep01,meena20}. 

Importantly, if the GW wavelength, $c/\fGW$, is comparable to or greater than the Schwarzschild radius, $2 G M_{L}/c^2$, of any given microlens, the wave will be diffracted by the gravitational `slit' \citep{nak99,tak03}. For $M_{L} \sim M_{\odot}$, this condition is fulfilled even for $\fGW \lesssim 10^2$ kHz. If $\fGW$ is too low relative to $M_{L}$ however, the resulting amplifications will be virtually invisible; $\sim$kHz band radiation lensed by stars thereby lies in something of a sweetspot. Geometric optics is inappropriate in this case, and the relevant mathematical object is instead the Fresnel-Kirchhoff diffraction integral \eqref{eq:diffint}. This integral is solved in this paper for a variety of microlens distributions (Figs. \ref{fig:grav25} and \ref{fig:grav250}) using the PL approach described by \cite{feld19,feld20a,feld20b}. 

Depending on the nature of the macrolens and the neutron star trajectory relative to it, we find that wave-optical lensing may warp the waveform to make it appear as though the ellipticity were varying (see Figs. \ref{fig:intensity25} and \ref{fig:intensity250}), which has implications when estimating the intrinsic properties of the neutron star through \eqref{eq:massquad} and similar formulae \citep{cio09,mast11,lander13}. Furthermore, the overall SNR during an observational run may increase for large amplification factors; see expression \eqref{eq:snr}. If continuous GWs emitted from a source $\sim$10 kpc away were lensed by (sections of) a cluster similar to that of 47 Tuc ($D_{\text{OL}} \sim 4$ kpc, $n \sim 10^5$), for example, we find that potentially large amplifications could be achieved {(possibly even larger than those seen in Fig. \ref{fig:intensity2d250})}. \cite{ras21} estimate, for 47 Tuc specifically, that the self-lensing rate for neutron stars is $\sim 2 \times 10^{-3} \text{ yr}^{-1}$. Any flux enhancement may however by offset by the loss of coherence due to phase modulation (Fig. \ref{fig:phase25}), which would inhibit matched filtering. %implying an inference of the stellar magnetic field would be off by a factor $\lesssim \sqrt{3}$.

Aside from GWs, wave-like effects are also important in the characterisation of radio sources in a variety of circumstances \citep{mun16,jow20,jow21}. For planetary-mass microlenses and $\lesssim$ GHz band sources, such as radio pulsars or fast radio bursts (FRBs), lensing is likely to operate in the diffractive regime where geometric optics fails to capture the salient features \citep{feld19}. The implementation of the core PL ideas is largely the same in this case, though for radio sources one typically requires a {larger} $\Delta \theta$ [by a factor $\propto (M_{L} / M_{\odot} ) (f_{\text{radio}}/\fGW)$] to adequately resolve the angular variations (see Sec. 4.1). An exploration of such cases for $n \gtrsim 10$ is currently beyond our available computational resources; see \cite{feld20b} for $n \leq 3$ simulations and a discussion on the relevant challenges faced when extending to $n > 3$. It is worth pointing out that the formalism presented here is not restricted to gravitational lensing, but can also be applied to the case of \emph{plasma} lensing, where a similar Fresnel-Kirchhoff integral arises; see also \cite{jow21} for recent applications of PL theory to plasma lensing.

We close by noting that there are multiple avenues for extension of this work. For the Galactic sources considered here, the lensing probability is at most a few by $10^{-6}$ for sources within the Galactic bulge \citep{pac86b,dep01}. This probability increases by several orders of magnitude for extragalactic sources, {where incoming radiation may be lensed by up to $\lesssim 10^{7}$ individual perturbers \citep{pac86}}. While continuous waves from sources at redshifts $z \gg 0$ will not be observable for the foreseeable future, burst signals from merger events are now routinely detected out to cosmological distances \cite[the record holder being GW190521 at a redshift $z = 0.82^{+0.28}_{-0.34}$;][]{ligo20}. Designing numerically efficient tools for the study of ever-higher $n$ simulations is thus of astrophysical importance \citep{guo20} {(see also Sec. 5.4)}. In any case, the results presented here are mostly intended as a proof-of-principle, and thus the lens-plane distributions we have adopted (Figs. \ref{fig:grav25} and \ref{fig:grav250}), while loosely resembling what might be expected of open or globular clusters, are also rather arbitrary. It would be interesting to instead model the microlenses based on astrophysical data. Finally, in an effort to better understand the nature of the GWs in strong gravitational fields, one might ambitiously try to extend the PL formalism beyond Newtonian backgrounds described by \eqref{eq:metric}; the lensing theory developed by \cite{hart19} and \cite{cus20} would be useful in this direction. More ambitiously still, wave-optical lensing in theories beyond general relativity could also be explored \citep{ezq20}.

\section*{Acknowledgements}
I am indebted to Mark Walker and Artem Tuntsov for introducing me to several key references about wave optics, microlensing, and Picard-Lefschetz theory, and for providing constructive criticism on an earlier version of this work. {I also thank the anonymous referee for providing helpful feedback, which improved the quality of this manuscript.}

%%%%%%%%%%%%%%%%%%%%%%%%%%%%%%%%%%%%%%%%%%%%%%%%%%%%%%%%%%%%%%%%%%%%%%%%%%%%%%%%%%%%%%%

%%%%%%%%%%%%%%%%%%%%%%%%%%%%%%%%%%%%%%%%%%%%%%%%%%%%%%%%%%%%%%%%%%%%

\appendix
\section{Picard-Lefschetz evaluation of the generalised Fresnel integral}

%where \eqref{eq:analyticsol} follows from results concerning the hypergeometric function.

In this Appendix, we outline how the PL calculation proceeds in a non-trivial case, possessing both singularities and images, where a closed-form solution is known. Consider the generalised Fresnel integral with \cite[e.g.,][]{math12},
\begin{align}
\mathcal{F}_{n,m} &= \int^{\infty}_{0} x^{m} e^{i x^{n}} dx  \label{eq:genfres} \\
&= \frac{1}{n} \Gamma \left( \frac {m+1} {n} \right) e^{\frac {i \pi (m+1)} {2n}}.  \label{eq:analyticsol}
\end{align}
Integrals of the form \eqref{eq:genfres} are similar to the Fresnel-Kirchhoff case of interest since we can write $x^{m} e^{i x^{n}} = e^{i x^n + m \log x}$, and the logarithmic term resembles the projection of the gravitational potential onto the lens plane, with $m$ playing the role of the lens mass; see equations \eqref{eq:timedelay} and \eqref{eq:psipot}. For a thorough discussion on the PL evaluation of the ordinary Fresnel integral ($m=0, n=2$), we refer the reader to \cite{feld19}.

%\hspace{-0.6cm}

To evaluate expression \eqref{eq:genfres} using a PL approach, we first analytically continue the variable, $x \rightarrow u + iv$, and expand the exponent into real and imaginary components. We consider the case of integer $n\geq2$ so that the integral converges and we can apply a Binomial expansion, ultimately finding
\begin{widetext}
\begin{equation} \label{eq:expd}
\begin{aligned}
\hspace{-0.31cm}i x^n + m\log x \rightarrow \sum_{n-k \text{ odd}}^{n} (-1)^{\tfrac{n-k+1}{2}} {n \choose k} u^{k} v^{n-k} + \frac{m}{2} \log ( u^2 + v^2)  + i \left[ \sum_{n-k \text{ even}}^{n} (-1)^{\tfrac{n-k}{2}} {n \choose k} u^{k} v^{n-k} + m \tan^{-1}\tfrac{v}{u} \right],
\end{aligned}
\end{equation}
\end{widetext}
where, for real $m$, the first group of terms (the function $h$ from Sec. 4) is strictly real and the second ($H$) is imaginary; for general $m \in \mathbb{C}$, the decomposition is only slightly more involved. The Morse flow equations \eqref{eq:morseflow} can now be written down in full, though are lengthy for general $n, m$. In the case of $n=4$, for example, they are explicitly given by
\begin{equation} \label{eq:ueqn}
\frac {d u} {d \lambda} = 12 u^2 v - 4 v^3 - \frac {m u} {u^2 + v^2},
\end{equation}
and
\begin{equation} \label{eq:veqn}
\frac {d v} {d \lambda} = 4 u^3 - 12 u v^2 - \frac {m v} {u^2 +v^2},
\end{equation}
where we use the Euclidean metric, $G^{ij} = \delta^{ij}$, by identifying $\mathbb{C}$ with $\mathbb{R}^2$ (see Footnote 3). For general $n,m$ it is not possible to evaluate the flow solutions analytically, owing largely to the non-linearity of the problem. A straightforward but numerical solution to \eqref{eq:ueqn}--\eqref{eq:veqn} is necessary. The initial conditions are set as $u(0)=v(0)=\epsilon=10^{-4}$ to circumvent the singularity at $x=0$ for $m \neq 0$. This regularisation is that which is outlined in point 2 within Sec. 4.1. Alternatively, we could start flowing from $u(0) = \epsilon, v(0) =0$ (dotpoint 3).
%Expansion \eqref{eq:expd} defines the functions $h$ and $H$ introduced in Sec. 4, respectively.

It is also not hard to see that there are, in general, stationary point(s) of the integrand \eqref{eq:genfres} at $p_{n,m} = (i m/n)^{1/n}$; at this (multi-valued) point, the right-hand-sides of \eqref{eq:ueqn} and \eqref{eq:veqn} vanish, and the flow terminates. Therefore, once these points are encountered, it is necessary to introduce another perturbation $u'(\lambda) \rightarrow u'(\lambda) + \epsilon$ to continue the flow (see Fig. \ref{fig:cont}). However, most of these images are irrelevant: those with $u < 0$, for example, lie outside of the domain. The overall thimble we require therefore consists of both upwards and downwards flows only from relevant images.

\begin{figure}[h]
\centering
  \includegraphics[width=0.445\textwidth]{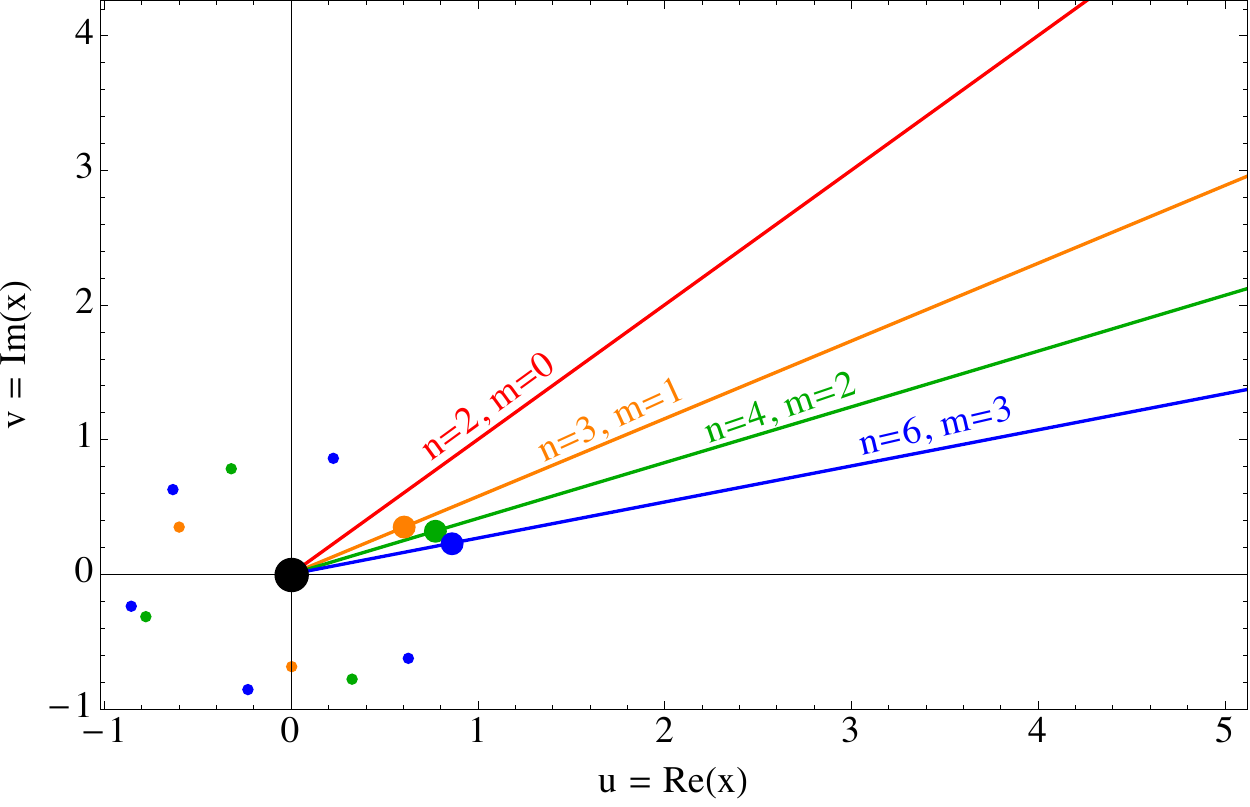}
  \caption{Examples of Lefschetz thimbles (solid curves) used in the evaluation of the generalised Fresnel integral, $\mathcal{F}_{n,m}$. The coloured points mark the locations of the $n$ images, $p_{n,m} = (i m/n)^{1/n}$, with the largest marker indicating the image that is relevant in the construction of the associated thimble, defined as a solution to the Morse flow equations \eqref{eq:morseflow}. In all cases with $m>0$, there is also a singularity at $x=0$, as shown by the black point. For $m=0$, this point is an image. Cauchy's theorem \eqref{eq:cauchy} is applied together with Jordan's lemma to evaluate the integral along $\text{Re}(x)$.
}
  \label{fig:lines}
\end{figure}

Solutions to the respective flow equations for various $m$ and $n$ are shown in Fig. \ref{fig:lines}. Note that these contours do not extend to infinity, and therefore the application of Cauchy's theorem is only approximate. However, even for these `short' thimbles, the errors between our results and the analytic solution \eqref{eq:analyticsol} are at the $\sim0.1\%$ level. For example, the thimbler yields a value $\text{LT}_{4,1} = 0.3128 + 0.3123 i$ while the true value, from \eqref{eq:analyticsol}, is $\mathcal{F}_{4,1} = 0.3133 + 0.3133 i$. Extending the range to beyond $\text{Re}(x) > 5$ reduces the error further; for thimbles built out to $\text{Re}(x) \gtrsim 10$, the numerical evaluations match the analytic expression to machine precision. %.

\section{Numerical convergence tests}

%$\Delta \theta$ variation, give a table.

In addition to the one-dimensional case considered in Appendix A, we can also examine how the PL evaluation depends on the angular resolution for two dimensional integrals. To this end, we consider the point-mass case (i.e., $n=1$), where the lens mass is placed at the origin without loss of generality. Ignoring offsets of the time-delay function, the amplification factor \eqref{eq:diffint} is given by the analytic formula [see equation (17) in \cite{tak03}]
\begin{equation} \label{eq:formula}
\begin{aligned}
F(\fGW,\boldsymbol{x}_{s}) =& \exp \left( \frac{w \pi} {4} + \frac {i w}{2} \ln\frac {w}{2} \right) \Gamma \left( 1 - \frac {i w} {2} \right) \\
&\times  {}_{1}F_{1} \left( \frac{ i w} {2}, 1; \frac {i w |\boldsymbol{x}_{s}|^2} {2} \right),
\end{aligned}
\end{equation}
where we use the shorthand $w = 8 \pi G M_{L} \fGW /c^3$, and ${}_{1}F_{1}$ denotes the hypergeometric function of the first kind.

In the PL integration, there are three separate `resolutions' that control the accuracy of the evaluation for any given source coordinates $\boldsymbol{x}_{s}$, the first of which is (i) the step size in the Morse solver. As commented in Sec. 4.1, we employ a Runge-Kutta algorithm to solve the Morse flow equation \eqref{eq:morseflow}, the step size of which is chosen sufficiently small such that the global error is at most one part in $\sim 10^{4}$, in turn ensuring that the imaginary part of the phase, defined by the function $H$, varies negligibly on the numerical thimble. The second source of error, related to the first, concerns (ii) the contour length (i.e., the maximum $\lambda$ out to which one flows). Formally the contour should extend to infinity to apply Cauchy's theorem, and thus a premature truncation necessarily implies approximation. However, the steepest descent nature of the Morse flow guarantees that this error is tiny even for short thimbles, as explored in Appendix A. Finally, we have (iii) the angular resolution, relevant when computing multi-dimensional integrals. The accuracy, as a function of this last resolution, depends on the numerical method used to sum the angular integrals -- Simpson's method is employed here.

Table \ref{tab:tab1} compares the numerical PL evaluation of the diffraction integral \eqref{eq:diffint} with $\fGW = 1$kHz, in the case of a single point-mass lens, with the analytic formula \eqref{eq:formula} for several different lens masses and positions over a variety of angular resolutions $2\pi / \Delta \theta$. In general, convergence for higher masses (or, equivalently, greater $\fGW$) requires greater $\Delta \theta$ because the integrand varies more rapidly. We see that for $\Delta \theta = 64$ the numerical expressions are relatively inaccurate, especially in the imaginary sector, with the exception of the lightest case $M_{L} = M_{\odot}$, where the results match the analytic expression to within $\sim 0.1\%$. For $\Delta \theta = 128$, the results for $M_{L} \leq 10 M_{\odot}$ match the analytic solutions to high precision, though greater resolution is needed for the highest mass cases. For $M_{L} =10^{3} M_{\odot}$ in fact, even doubling the resolution again to $\Delta \theta = 256$ produces unreliable results. In that case, a resolution of $\Delta \theta = 512$ yields a match to within $\sim 0.2 \%$. It should be noted though that for $M_{L} \gtrsim 10^{3} M_{\odot}$ (i.e., $w \gg 1$) we are no longer in the diffractive regime, and a geometric optics approach would already be sufficient.

The heaviest macrolens considered in this paper has $\sum_{k} M_{k} = 250 M_{\odot}$ (i.e., $n=250$), and thus we anticipate that $\Delta \theta = 256$ yields maximum errors of a few percent for $\fGW \lesssim 2$kHz. For all simulations presented in this work, we use $\Delta \theta = 512$.

%Given

%In those cases, we find 

%We find that $\Delta \theta = 256$ yields 

% the numerical results  Furthermore, even at half this resolution, $\Delta \theta = 32$, 

%In the case of a point mass lens, the mass and source frequency are degenerate, as the analytic formula depends on their product. It is because of this that the convergence of the numerical thimbler is marginally worse in the 2kHz case for a given $\Delta \theta$. , larger $\Delta \theta$ values by a factor $\sim 4$ are necessary to achieve the same levels of accuracy. 

\begin{table}[h]
\caption{Comparison of numerical evaluations of the amplification factor $F$ for a point mass lens located at the origin with the analytic formula \eqref{eq:formula}. In each case, the contour length and Morse solver step-size is held fixed, as is the radiation frequency, $\fGW = 1$kHz.}
\hspace{-1cm}\begin{tabular}{c  c  c  c} 
%\hline \hline
 $\left(|\boldsymbol{x}_{s}|, M_{L} \right)$ & $\Delta \theta$ & Numerical & Exact  \\
 \hline
 (0, $M_{\odot}$) & 64 & 1.08907 - 0.14960 $i$ & 1.0883 - 0.14942 $i$ \\
 & 128 & 1.0885 - 0.14946 $i$ &  \\
 & 256 & 1.0883 - 0.14942 $i$ &  \\
 \hline
 (5, $M_{\odot}$) & 64 & 0.99437 - 0.17269 $i$ & 0.99442 - 0.17238 $i$ \\
 & 128 & 0.99440 - 0.17245 $i$ &  \\
 & 256 & 0.99442 - 0.17238 $i$ &  \\
 \hline
 (0, $10 M_{\odot}$) & 64 & 1.9981 - 0.04398 $i$ & 1.9905 - 0.04087 $i$ \\
 & 128 & 1.9924 - 0.04163 $i$ &  \\
 & 256 & 1.9909 - 0.04104 $i$ &  \\
  \hline
 (5, $10 M_{\odot}$) & 64 & -0.46442 - 0.99317 $i$ & -0.42471 - 0.94145 $i$ \\
 & 128 & -0.42474 - 0.94147 $i$ &  \\
 & 256 & -0.42471 - 0.94145 $i$ &  \\
  \hline
 (0, $10^2 M_{\odot}$) & 64 & 4.0475 - 4.8647 $i$ & 3.9867 - 4.7883 $i$ \\
 & 128 & 3.9964 - 4.8010 $i$ &  \\
 & 256 & 3.9961 - 4.7902 $i$ &  \\
 \hline
 (5, $10^2 M_{\odot}$) & 64 & 0.0349 - 0.76385 $i$ & 0.25649 - 0.95928 $i$ \\
 & 128 & 0.30569 - 1.0736 $i$ &  \\
 & 256 & 0.25319 - 0.95975 $i$ &  \\
 \hline
 (0, $10^3 M_{\odot}$) & 128 & -6.3551 - 19.680 $i$ & -5.0514 - 19.045 $i$ \\
 & 256 & -5.3672 - 19.200 $i$ &  \\
 & 512 & -5.0464 - 19.042 $i$ &  \\
 \hline
 (5, $10^3 M_{\odot}$) & 128 & 0.84169 - 0.05703 $i$ & 0.97268 - 0.16039 $i$ \\
 & 256 & 0.85180 - 0.13543 $i$ &  \\
 & 512 & 0.97147 - 0.16098 $i$ &  \\
 \hline
 \label{tab:tab1}
\end{tabular}
\end{table}

%Need to comment on there being hidden parameters in the code also (i) time stepper in morse solver, (ii) contour length (see App A), and (iii) delta 

%As is evident in the above, including higher masses is the same as increasing the frequency.

\end{document}